\documentclass[traditabstract]{aa}
\pdfoutput = 1
\usepackage{graphicx}
\usepackage{epstopdf}
\usepackage{amsmath}
\usepackage{amssymb}
\usepackage{deluxetable}
\usepackage{longtable}
\usepackage{lscape}
\usepackage{rotating}
\usepackage{natbib}
\usepackage{epsfig}
\usepackage{array}
\newcommand{\lsim}{\lower 2pt \hbox{$\, \buildrel {\scriptstyle<}\over {\scriptstyle \sim}\,$}}  
\newcommand{\gsim}{\lower 2pt\hbox{$\, \buildrel {\scriptstyle >}\over {\scriptstyle \sim}\,$}}

\newcommand{\oxythr}{{[OIII]}}

\newcommand{\kms}{km s$^{-1}$ }

\newcommand{\lala}{$\lambda\lambda$}

\bibpunct{(}{)}{;}{a}{}{,} 

\begin{document}

\title{Extended Warm Gas in the ULIRG Mrk273: Galactic Outflows and Tidal Debris}

\author{{\small J. Rodr\'iguez Zaur\'in\inst{1,2}, C. N. Tadhunter\inst{3},
    D. S. N. Rupke\inst{4}, S. Veilleux\inst{5}, H.  W. W. Spoon\inst{6}, M.
    Chiaberge\inst{7}, C. Ramos Almeida\inst{1,2}, D. Batcheldor\inst{8} and
    W. B. Sparks\inst{7}}}

\offprints{Javier Rodr\'iguez Zaur\'in\\ {\tt javierr@iac.es} \smallskip}

\institute{Instituto de Astrof\'isica de Canarias, E-38205, La Laguna, Tenerife,
  Spain \\email: {javierr@iac.es} 
\and Departamento de Astrof\'isica, Universidad de La Laguna, Tenerife, Spain
\and Department of Physics and Astronomy, University of Sheffield, Sheffield S3
7RH
\and Department of Physics, Rhodes College, Memphis, TN 38112, USA 
\and Department of Astronomy, University of Maryland, College Park, MD 20742,
USA
\and Cornell University, CRSR, Space Sciences Building, Ithaca, NY 14853, USA
\and Space Telescope Science Institute, 3700 San Martin Drive, Baltimore, MD 21218, USA
\and Physics and Space Sciences Department, Florida Institute of Technology, 150
West University Boulevard, Melbourne, FL 32901, USA
}

\date{  }

\abstract{We present new {\it HST} ACS medium- and narrow-band images and
  long-slit, optical (4000 - 7200\AA) spectra obtained using the Isaac Newton
  Telescope (INT) on La Palma, of the merging system Mrk273. The {\it HST}
  observations sample the [OIII]$\lambda\lambda$4959,5007 emission from the
  galaxy and the nearby continuum. These data were taken as a part of a larger
  study of ultraluminous infrared galaxies (ULIRGs) with the aim of
  investigating the importance of the warm, AGN induced outflows in such
  objects. The {\it HST} images show that the morphologies of the extended
  continuum and the ionized gas emission from the galaxy are decoupled,
  extending almost perpendicular to each other. In particular, we detect for the
  first time a spectacular structure of ionized gas in the form of filaments and
  clumps extending $\sim$23kpc to the east of the nuclear region. The quiescent
  ionized gas kinematics at these locations suggests that these filaments are
  tidal debris left over from a secondary merger event that are
  illuminated by an AGN in the nuclear regions. The images also reveal a complex
  morphology in the nuclear region of the galaxy for both the continuum and the
  [OIII] emission. Consistent with this complexity, we find a wide diversity of
  emission line profiles in these regions. Kinematic disturbance, in the form of
  broad (FWHM $>$ 500 km s$^{-1}$) and/or strongly shifted ($|\Delta$V$|$ $>$150
  km s$^{-1}$ ) emission line components, is found at almost all locations in
  the nuclear regions, but confined to a radius of $\sim$4 kpc to the east and
  west of the northern nucleus. In most cases, we are able to fit the profiles
  of all the emission lines of different ionization with a kinematic model using
  2 or 3 Gaussian components. From these fits we derive diagnostic line ratios
  that are used to investigate the ionization mechanisms at the different
  locations in the galaxy. We show that, in general, the line ratios are
  consistent with photoionization by an AGN as the main ionization
  mechanism. {Finally, the highest surface brightness [OIII] emission is found
    in a compact region that is coincident with the so-called SE nuclear
    component.} The compactness, kinematics and emission line ratios of this
  component suggest that it is a separate nucleus with its own AGN. At this
  stage, further observations are required to confirm the dual (or multiple?)
  AGN nature of Mrk273.}

\keywords{Galaxies: evolution -- galaxies: starburst -- galaxies:active.}

\authorrunning{J. Rodr\'iguez Zaur\'in et al.:}
\titlerunning{Outflows and tidal debris in Mrk273}

\maketitle 
%
\section{Introduction}

Fast outflows induced by the central QSO are now almost invariably detected in
dusty mergers with AGN nuclei
\citep[e.g.][]{Westmoquette12,Rodriguez-Zaurin13,Veilleux13,Rupke13b}. These
outflows are a key element in simulations of galaxy evolution through major
mergers, where they are used to regulate the correlation between the black hole
mass and the host galaxy properties
\citep[e.g.][]{Silk98,Fabian99,Benson03,Johansson09}. In particular, they are
predicted to be extremely powerful towards the final stages of the merger, when
the merging nuclei coalesce, potentially quenching the surrounding star
formation activity
\citep[e.g.][]{diMatteo05,Springel05,Johansson09,Hopkins10}. However, from an
observational perspective, the nature of their interaction with the ISM in the
host galaxies and importance relative to the outflows induced by star formation
activity, have yet to be firmly established.

In this context Ultra Luminous Infrared Galaxies (ULIRGs, L$_{\rm IR} >$
10$^{12}$ L$_{\odot}$) are ideal objects to investigate the physical processes
involved in AGN-induced outflows. Their prodigious far-IR radiation represents
the dust re-processed light of nuclear power sources (starbursts and/or AGN). In
addition, ULIRGs almost invariably show morphological evidence consistent with
triggering of the activity in major galaxy mergers
\cite[e.g.][]{Murphy96,Clements96,Sanders96,
  Surace00a,Surace00b,Scoville00,Veilleux02}. Therefore, ULIRGs represent just
the situation modelled in many of the most recent hydrodynamic simulations.

As expected from the merger simulations, AGN-induced outflows in ULIRGs have
recently been found in all gas phases: neutral \citep{Rupke11,Rupke13b}, ionized
\citep[e.g.][]{Westmoquette12,Rodriguez-Zaurin13,Rupke13b} and molecular
\citep[e.g.][]{Fischer10,Feruglio10,Sturm11,Veilleux13,Cicone12,Cicone13,
  Rupke13a,Spoon13}. For the ionized and neutral gas, the outflows in ULIRGs
with AGN show velocity widths (FWHM) and shifts of up to 1700 km s$^{-1}$ and
2000 km s$^{-1}$ respectively\footnote{Note that throughout the paper, when we
  refer to the outflow velocity shifts from the work of Rupke and collaborators,
  we specifically mean the $v_{50\%}$ values as defined in their papers.}
\citep[][hereafter RZ13 and RV13]{Rodriguez-Zaurin13,Rupke13b}. These numbers
are substantially smaller for ULIRGs with no detected AGN activity (FWHM
typically less than 500 km s$^{-1}$, RV13, RZ13). In the molecular phase, the CO
emission and/or OH absorption lines show velocity shifts of up to 1000 km/s in
some ULIRGs with AGN activity \citep[][]{Sturm11,Veilleux13,Cicone13}. As in the
cases of the ionized and neutral phases, the velocity shifts are smaller for
those ULIRGs powered by starburst \citep[e.g.][]{Veilleux13,Cicone13}.

Recent integral field spectroscopic (IFS) studies of local samples of ULIRGs
have been used to investigate not only the kinematics, but also the spatial
structures and the geometries of these outflows
\citep[e.g.][RV13]{Colina99,Bedregal09,Westmoquette12}. For example, RV13 found
a variety of outflow morphologies, from collimated bipolar winds, sometimes in
the form of ``super-bubbles'', to less collimated shocks extendeding on
kiloparsecs scales in the galaxies. However, these studies are usually
concentrated in the central few kiloparsecs of the objects, and therefore, it is
possible that they do not sample the full extent of the outflowing material. In
addition, IFS data are usually limited to a relatively narrow wavelength range,
making it hard to determine the ionization mechanism responsible for the
outflows. Due to these limitations, the mass outflow rates and kinetic powers
derived from these studies remain uncertain.

As one of the closest ULIRGs, Mrk273 (z = 0.0373, L$_{\rm IR}$ = 10$^{12.21}$
L$_{\odot}$) represents a key target for studies aimed at understanding the
nature of ULIRGs and their associated outflows.  This late-merger system shows
an impressive tidal tail extending over 40 kpc to the south of the nuclear
region. It has been studied extensively at all wavelengths, from the UV to the
X-rays
\citep[e.g.][]{Condon91,Soifer00,Armus07,Howell10,Rodriguez-Zaurin10,Iwasawa11,U13}. At
near- and mid-IR wavelengths, two nuclei become apparent, commonly referred to
as northern (N) and south-western (SW) components
\citep{Majewski93,Knapen97,Scoville00,Soifer00}. An additional third nuclear
structure, referred to as south eastern (SE) component, emerges at radio
wavelengths \citep[][]{Condon91}. The N nucleus is the strongest radio source
\citep[][]{Carilli00,Bondi05} and makes the main contribution to the MIR
emission \citep[][]{Soifer00}. Radio and CO(2--1) observations suggest that the
N nucleus is the site of the dense molecular gas disk of 1 $\times$ 10$^{9}$
M$_{\odot}$ containing a compact, powerful starburst formed by several luminous
supernovae or supernovae remnants \citep{Condon91,Carilli00,Bondi05}. However,
based on their recent AO IFS near-IR observations of the sources, \cite{U13}
found evidence for a super-massive black hole of mass (1.04$\pm$0.1)x10$^9$
solar masses in the N nucleus. In addition, the detection of enhanced Fe K
emission at the same location, suggests that this nucleus might contain a
heavily obscured AGN \citep{Iwasawa11}.

The SW component coincides with the location of the hard X-ray point source, and
is identified as the host of the AGN \citep{Iwasawa11}. On the other hand,
although the SE component was first identified as a star cluster based on NICMOS
images, the strength of the [Si{\small VI}] at this location suggests that this
region is at least partially ionized by an AGN, with the ionizing AGN perhaps
located in the N or SW nucleus \citep{U13}.

The dominant power source responsible for the IR luminosity of Mrk273 remains
uncertain. While the strong [O {\small IV}] and [Ne {\small V}] emission
relative to [Ne {\small II}] suggest heating by an AGN as the dominant mechanism
\citep{Genzel98,Veilleux09}, the L(MIR)/L(FIR) ratio is consistent with
starburst activity powering almost the entire IR luminosity of the source
\citep{Veilleux09}. In addition, other diagnostics based on the equivalent width
of the PAH feature at 7.7$\mu$m suggest that the contributions to the heating of
the dust by the AGN and starburst components are actually similar
\citep{Veilleux09}.

The nuclear kinematics of the source have been investigated in the past using
IFS techniques \citep[][RV13]{Colina99}. The RV13 study of the central 4.5
$\times$ 6 kpc revealed the presence an ``extended'' outflow plus a bipolar
superbubble, along with a rotating component that traces the CO(1--0) rotation
curve of \cite{Downes98}. The superbubble is aligned N-S, with projected
velocities of up to -1500 km s$^{-1}$, and is likely to be ionized by an AGN.

Although much attention has been paid to the nuclear regions of Mrk273, little
work has been done on the morphology, kinematics and ionization of the gas at
larger scales. In this paper we present new imaging and spectroscopic
observations of Mrk273 taken with the Hubble Space Telescope ({\it HST}) and the
Isaac Newton Telescope ({\it INT}). We use these combined datasets to relate the
properties of the ionized gas emission in the nuclear regions to those on more
extended scales. In this way we aim to establish whether the warm, AGN-driven
outflows are important for the overall evolution of the host galaxy. Throughout
the paper we adopt H$_{0}$ = 68 km s$^{-1}$, $\Omega_{\rm m}$ = 0.29 and
$\Omega_{\Lambda}$ = 0.71. At the redshift of the source, this gives a
luminosity distance of D$_{\rm L}$ = 171.5 Mpc and a scale of 0.762 kpc
arcsec$^{-1}$.

\section{Observations and data reduction}

\subsection{Imaging data}

\subsubsection{{\it HST} data}

\begin{figure*}
\centering
\begin{tabular}{cc}
\hspace{-6cm}\includegraphics[width=0.645\textwidth]{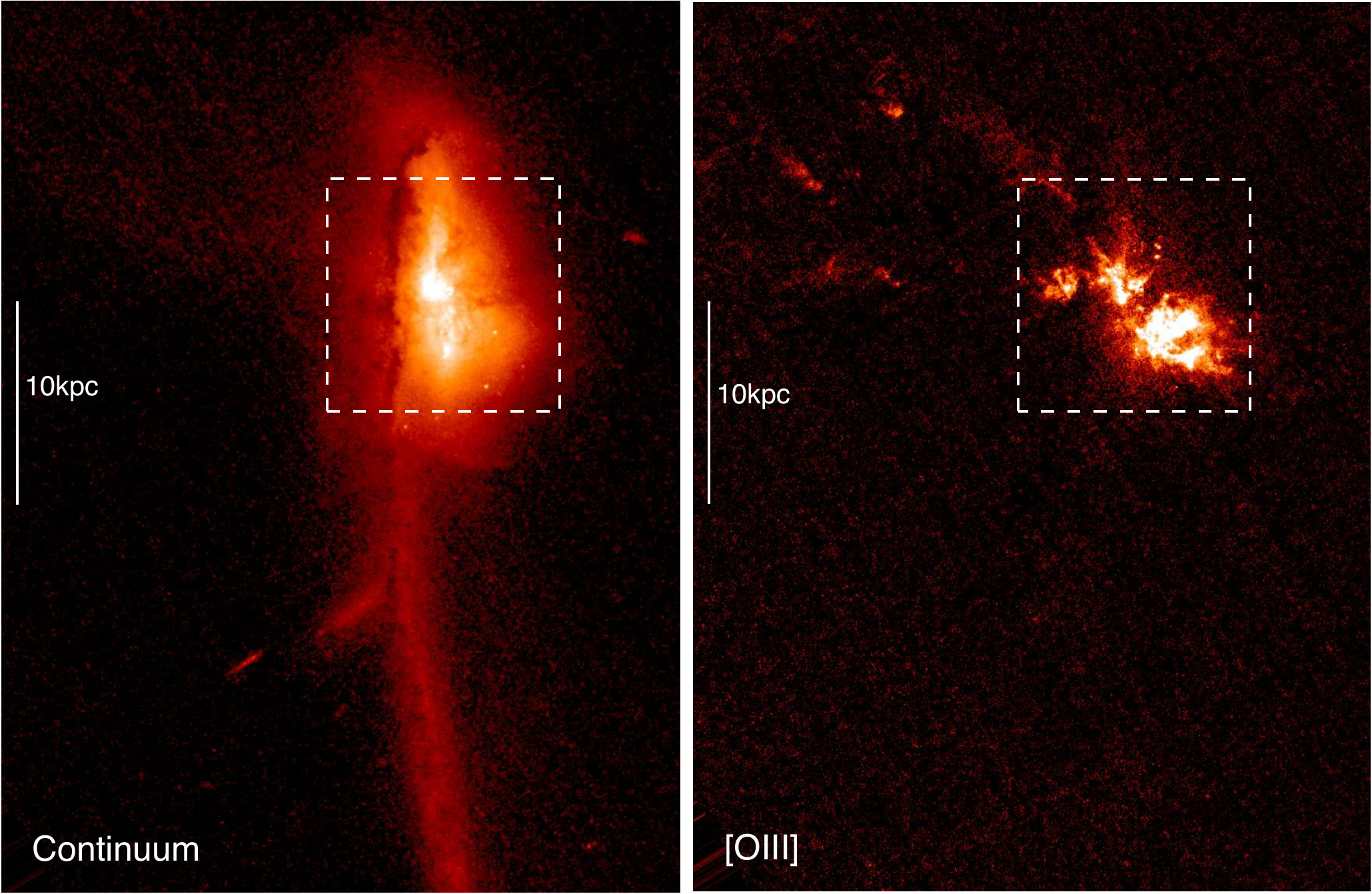}&
\hspace{-6.6cm}\includegraphics[width=0.32\textwidth]{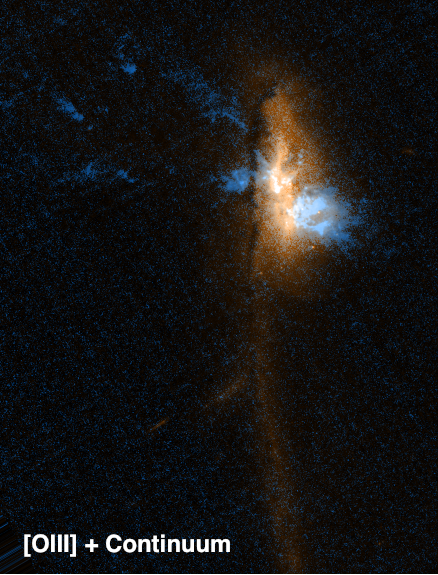}\\
\includegraphics[width=1.0\textwidth]{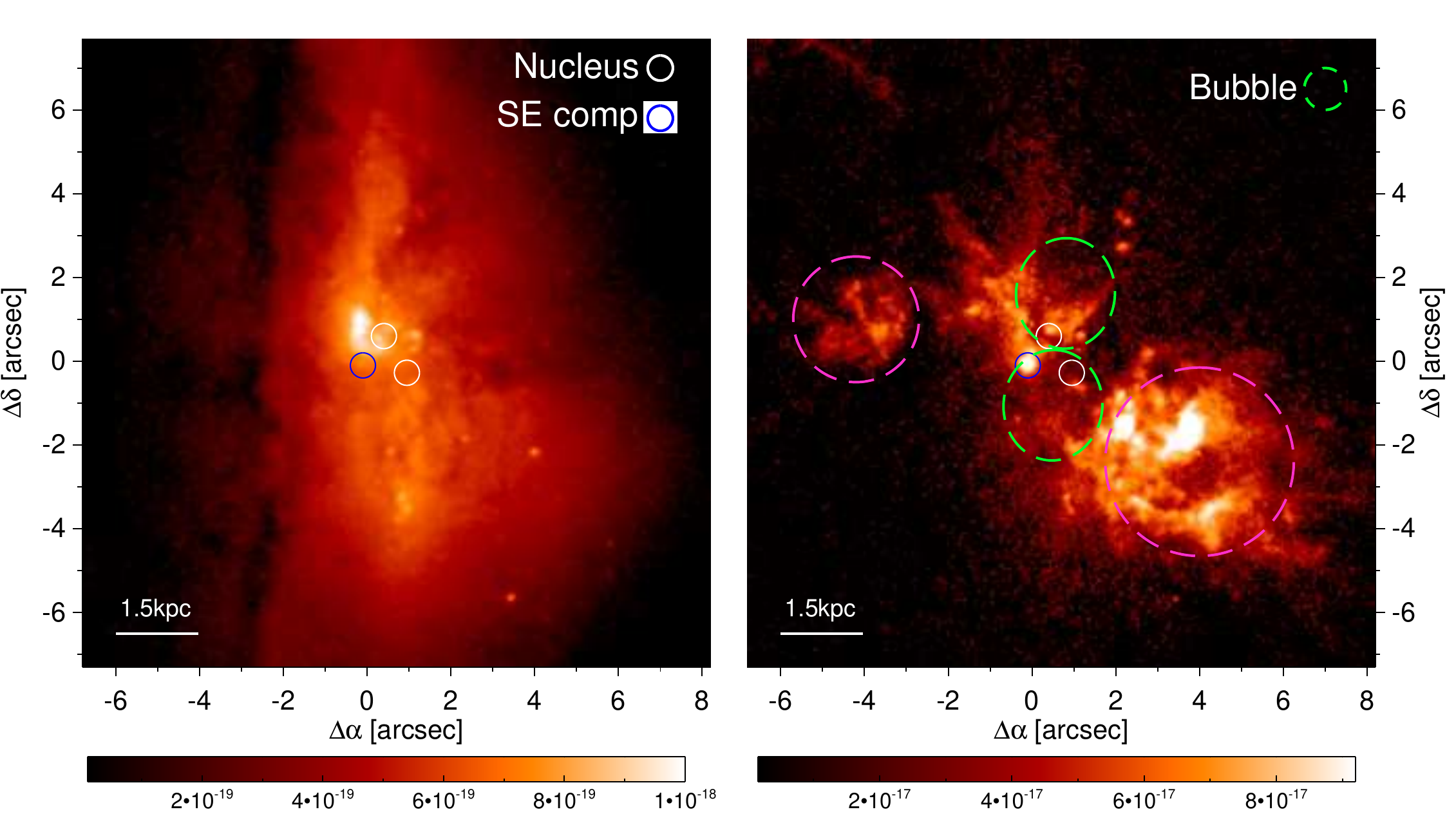}
\end{tabular}
\caption{Upper panel, left to right: {\it HST}-ACS continuum, continuum
  subtracted [OIII], and [OIII] (blue)+continuum (red) images of the
  galaxy. North is up, East is left. The dashed-line boxes indicated the zoomed
  region in the lower panel. Lower panel: zoom into the central 15$\times$15
  arcsec region (11.6$\times$11.6 kpc) for the continuum (upper-left) and the
  continuum-subtracted [OIII] (upper-middle) images of the galaxy. The color
  bars at the bottom of the images indicate the flux in units of erg s$^{-1}$
  cm$^{-1}$ \AA$^{-1}$ and erg s$^{-1}$ cm$^{-1}$ for the continuum and the
  [OIII] images respectively. The white circles indicate the locations of the N
  and SW nuclear components observed at IR wavelengths, while the blue circle
  corresponds to the location of the SE component that emerges at radio
  wavelengths. The green-dashed line ellipses in the continuum subtracted [OIII]
  image indicate the location of the nuclear superbubble reported by RV13 in
  their IFS study of the source, while magenta-dashed line circles indicate the
  location of the so-called Outflow-East and West (see text for details). A
  color version of this Figure is available in the online journal.}
\label{HST_frames}
\end{figure*}

New {\it HST} ACS images of Mrk273 were taken during Cycle 20 as a part of a
larger program aimed at studying the AGN-induced outflows in a sample of local
ULIRGs (GO:12934, PI:C.N. Tadhunter). The Wide Field Channel (WFC, 0.049 arcsec
pixel$^{-1}$) on the ACS was used in combination with the FR505N narrow-band
ramp filter, and the F550M medium-band filter. The former image was centered on
the redshifted [OIII]$\lambda$5007 (hereafter \oxythr) emission line, while the
F550M medium-band filter was centered on the nearby continuum towards redder
wavelengths. Details of the imaging observations are shown in Table
\ref{Im-table}.

\begin{table}
\begin{tabular}{llllll}
\hline\hline
Telescope & Camera & Filter & $\lambda_{c}$ & $\Delta\lambda$ & Exp time \\
&&&(\AA)&(\AA)& (s)\\
\hline
{\it HST} & ACS/WFC & F550M & 5581.5 &  384 & 716\\
&&FR505N & 5175.6 & 92 & 1356\\
{\it INT} & WFC & \#228 & 6813 & 93 & 2800 \\ 
\hline
\end{tabular}
\caption{Log of the imaging observations of Mrk273 (z = 0.0373) used for this
  paper.}
\label{Im-table}
\end{table}

The data were reduced using the standard data reduction pipeline procedures
which employ two packages: the {\sc CALACS} package, which includes dark
subtraction, bias subtraction and flat-field correction and produces calibrated
images, and the {\sc MULTIDRIZZLE} package, which corrects for distortion and
performs cosmic ray rejection. Any remaining cosmic rays were removed manually
using the routines {\sc IMEDIT} in {\sc IRAF} and/or {\sc CLEAN} within the {\sc
  STARLINK} package {\sc FIGARO}.

To convert into physical units we used the PHOTFLAM header keyword. PHOTFLAM is
the sensitivity conversion factor and is defined as the mean flux density
F$_{\lambda}$ (in units of erg cm$^{-2}$ \AA$^{-1}$ counts$^{-1}$) that produces
1 count per second for a given {\it HST} observing mode. Since drizzled ACS
images are in units of counts s$^{-1}$, these may simply be multiplied by the
PHOTFLAM value to obtain the flux in units of erg cm$^{-2}$ sec$^{-1}$
\AA$^{-1}$. The error associated with the flux calibration, including both the
photon noise and calibration uncertainty, is $\sim$5\% for both the continuum and
the [OIII] image.

At this stage it is important to add a caveat about the use of the PHOTFLAM
parameter. While PHOTFLAM is adequate when the flux is approximately constant
through the bandpass, it might not be adequate when the spectrum of the source
shows strong emission lines within the bandwidth, which is the case of our
\oxythr~image. Therefore, to check the accuracy of the calibartion of the {\it
  HST} images, we compared them with the \cite{Rodriguez-Zaurin09} WHT-ISIS long
slit spectroscopic observations of the source.

To that aim, it is crucial to know the precise location of the slit during the
observations. We first convolved the {\it HST} images with a Gaussian profile to
simulate the seeing conditions during the spectrocopic observations. Spatial
profiles of width identical to the slit-width of the spectra (1.5 arcsec) were
then extracted from the images, in steps of 2 pixels (0.1 arcsec). These spatial
profiles were compared with those of spatial slices extracted from the spectra
using wavelength ranges selected to match those of the {\it HST} filters, until
a match was found. Following this procedure we were able to find the location of
the slits on the image with a precision of 5 {\it HST} pixels (0.25
arcsec). Once the location of the slit was known, we integrated the emission
over the spatial extent of the galaxy covered by the slit for both the {\it HST}
and the {\it WHT} spatial profiles described before.

The results obtained from this procedure are two flux measurements that can be
directly compared to check the consistency between the flux calibrations. If we
refer to these flux values as F$_{\rm HST}$ and F$_{\rm WHT}$, we find that
F$_{\rm HST}$/F$_{\rm WHT}$ is equal to 1.05 and 1.10 for the continuum and
[OIII] observations respectively. Given the uncertainties associated with the
process (e.g. the varying seeing during the spectroscopic observations, or the 5
pixel uncertainty in the slit position) we find that the flux calibrations for
the {\it HST} and the {\it WHT} observations are in excellent agreement.

Once the images were calibrated in flux we subtracted the continuum from the
\oxythr~image to end up with an image that traces ``pure''
\oxythr~emission. Prior to this task, it was necessary to align the images. To
that aim we used a series of {\sc IRAF} routines. In the first place we used the
task {\sc SREGISTER} which registers an image to a reference image using
celestial coordinate information in the headers. In addition, there are three
foreground stars in the ACS/WFC field of view used for the Mrk273
observations. Therefore, it was possible to use the tasks {\sc GEOMAP} and {\sc
  GEOTRAN} to geometrically align the images and refine the final result. The
accuracy of the alignment was measured using the routine {\sc CENTER} that
calculates the centroids of the stars in the aligned images. We found that the
images are aligned with an accuracy better than 0.02 pixels. The continuum,
continuum subtracted, and \oxythr+continuum images are shown in Figure
\ref{HST_frames}.

\begin{figure}
\centering
\includegraphics[width=0.5\textwidth]{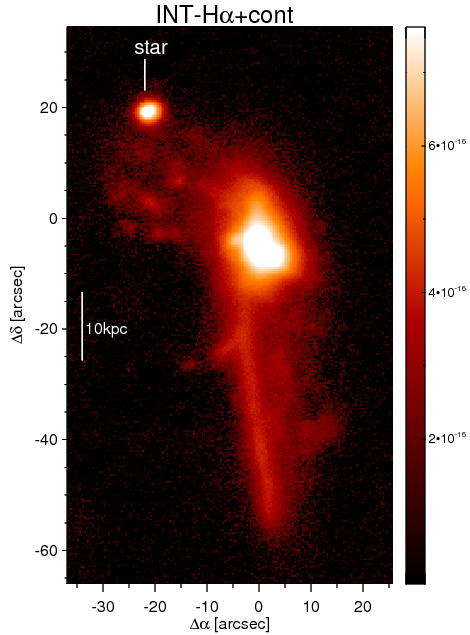}
\caption{{\it INT}-WFC H$\alpha$+[NII] image of Mrk273. No continuum emission
  was subtracted for this image. The color bar to the right of the image
  indicates the flux in units of ergs s$^{-1}$ cm$^{-2}$. The bright point
  source to the North-East of the galaxy is a star (a color version of this
  Figure is available in the online journal). }
\label{INT-WFC}
\end{figure}

\subsubsection{{\it INT}-WFC data}

A deep image of Mrk273 was taken in April 2013 using the Wide Field Camera (WFC)
mounted on the 2.5m {\it INT} telescope at the Observatorio del Roque de los
Muchachos (ORM), La Palma. The WFC is an optical mosaic camera consisting of 4
CCDs of 2098 $\times$ 4100 pixels (0.33 arcsec/pixel). Mrk273 was always
observed with CCD4, which has the highest efficiency at red wavelengths. For
these observations we used the Filter\#228 ($\lambda_{c}=6813$\AA, FWHM =
93\AA), which is centered on the H$\alpha$ emission line\footnote{We assume a
  systemic redshift of $z = 0.0373$ for the system (see RV13). All the radial
  velocity shifts presented in this paper are measured relative to this
  redshift.} and includes the [NII]$\lambda\lambda$6549,6583 doublet. In order
to remove the cosmic rays, hot pixels and other artifacts we took four 700s
exposures dithered on four positions in the sky. Unfortunately, due to some
technical difficulties during the night of the observations we could not acquire
a line-free continuum image in a wavelength range adjacent to H$\alpha$.
Therefore, no continuum subtraction could be performed in the case of the {\it
  INT}-WFC image of Mrk273 presented in this paper. The seeing during the
observations, measured with foreground stars in the images, was FWHM = 1.5 --
1.6 arcsec. Details of the INT/WFC observations are shown in Table
\ref{Im-table}

We used a series of {\sc IRAF} routines to perform the standard reduction
process, including bias subtraction, flat-field correction and cosmic ray
rejection. Any remaining cosmic rays after the initial reduction were cleaned
``manually'' using the routines {\sc IMEDIT} in {\sc IRAF} and/or {\sc CLEAN}
within the {\sc STARLINK} package {\sc FIGARO}. To transform into physical units
we used two standard stars observed with the exact same set-up used for the
observations of the galaxy, and calculate a factor to transform counts s$^{-1}$
into ergs s$^{-1}$ cm$^{-2}$. The calibration factor obtained for each of the
stars is identical and the estimated flux calibration uncertainty is 7\%. The
flux calibrated INT image is shown in Figure \ref{INT-WFC}.


Finally, we have followed the same procedure described in the previous section
for the {\it HST} images and compared the flux calibration of the {\it INT}-WFC
image with that of the existing {\it WHT}-ISIS spectroscopic observations of the
galaxy \citep{Rodriguez-Zaurin09}. If we define F$_{\rm WFC}$ as the integrated
flux over the spatial extent of the galaxy covered by the slit in the WFC image,
we find that F$_{\rm WFC}$/F$_{\rm WHT}$ = 0.8, i.e. the flux calibrations for
the two observations are in good agreement.

\begin{figure}
\centering
\includegraphics[width=0.45\textwidth]{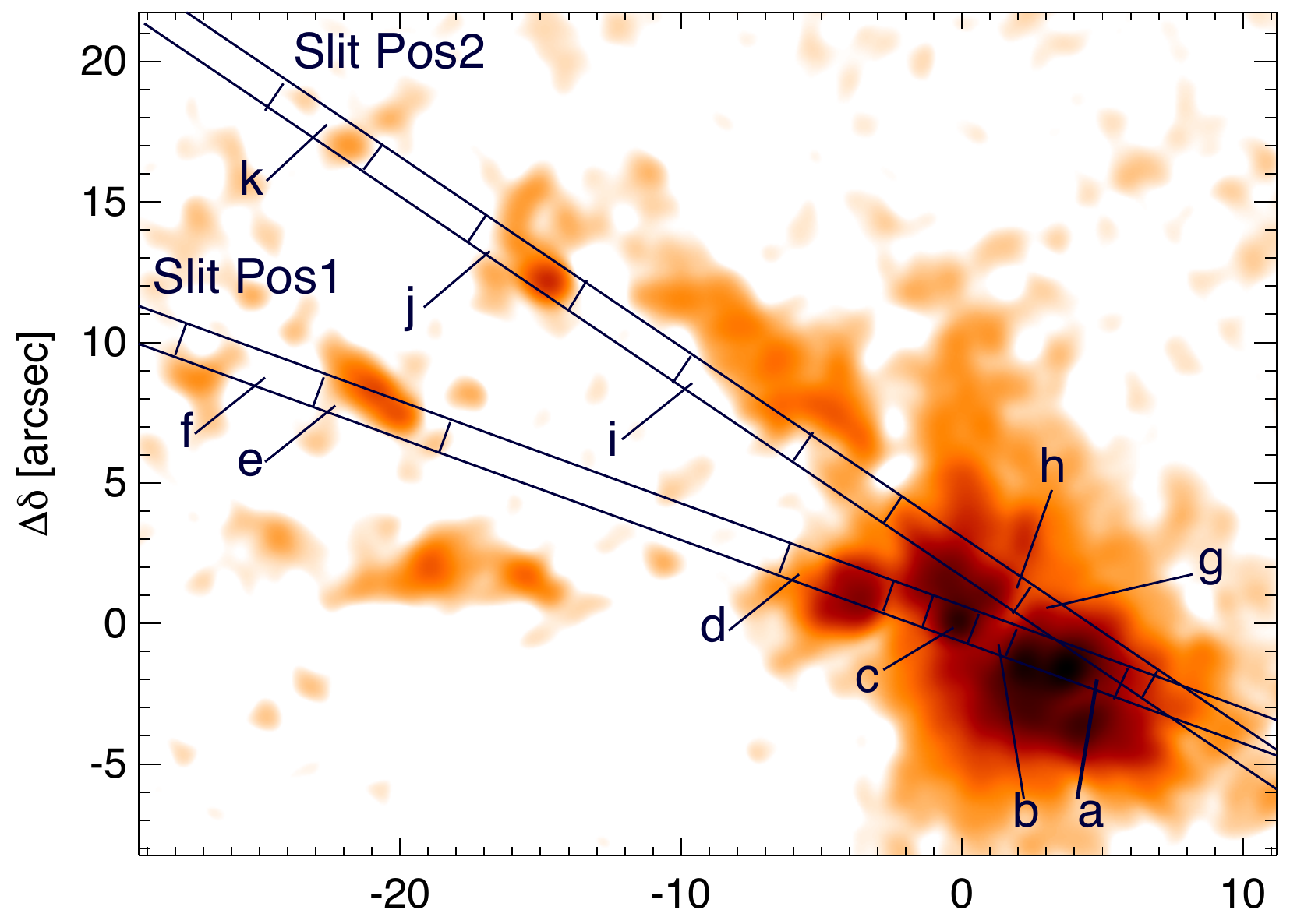}\\
\includegraphics[width=0.45\textwidth]{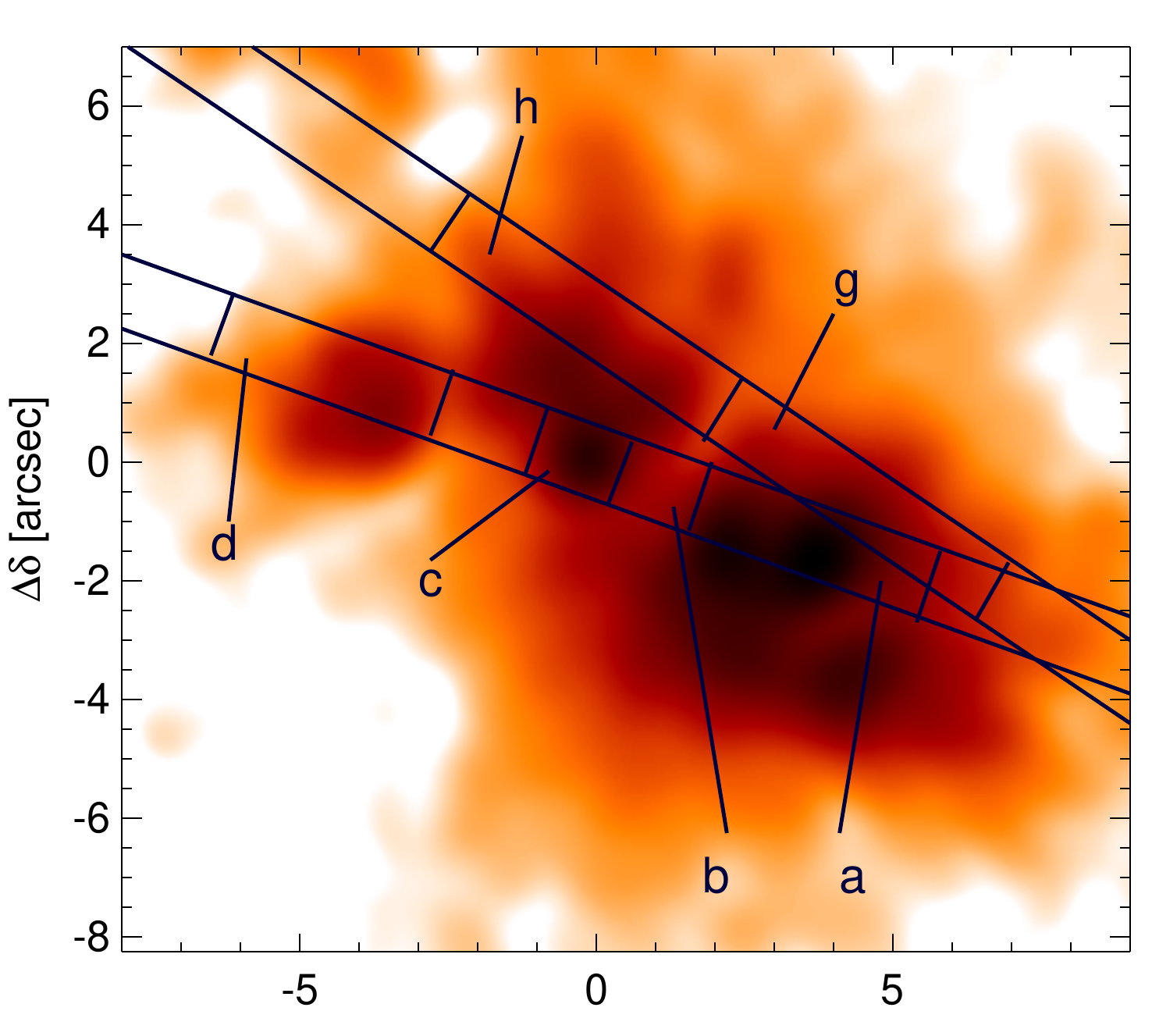}\\
\hspace{0.2cm}\includegraphics[width=0.46\textwidth]{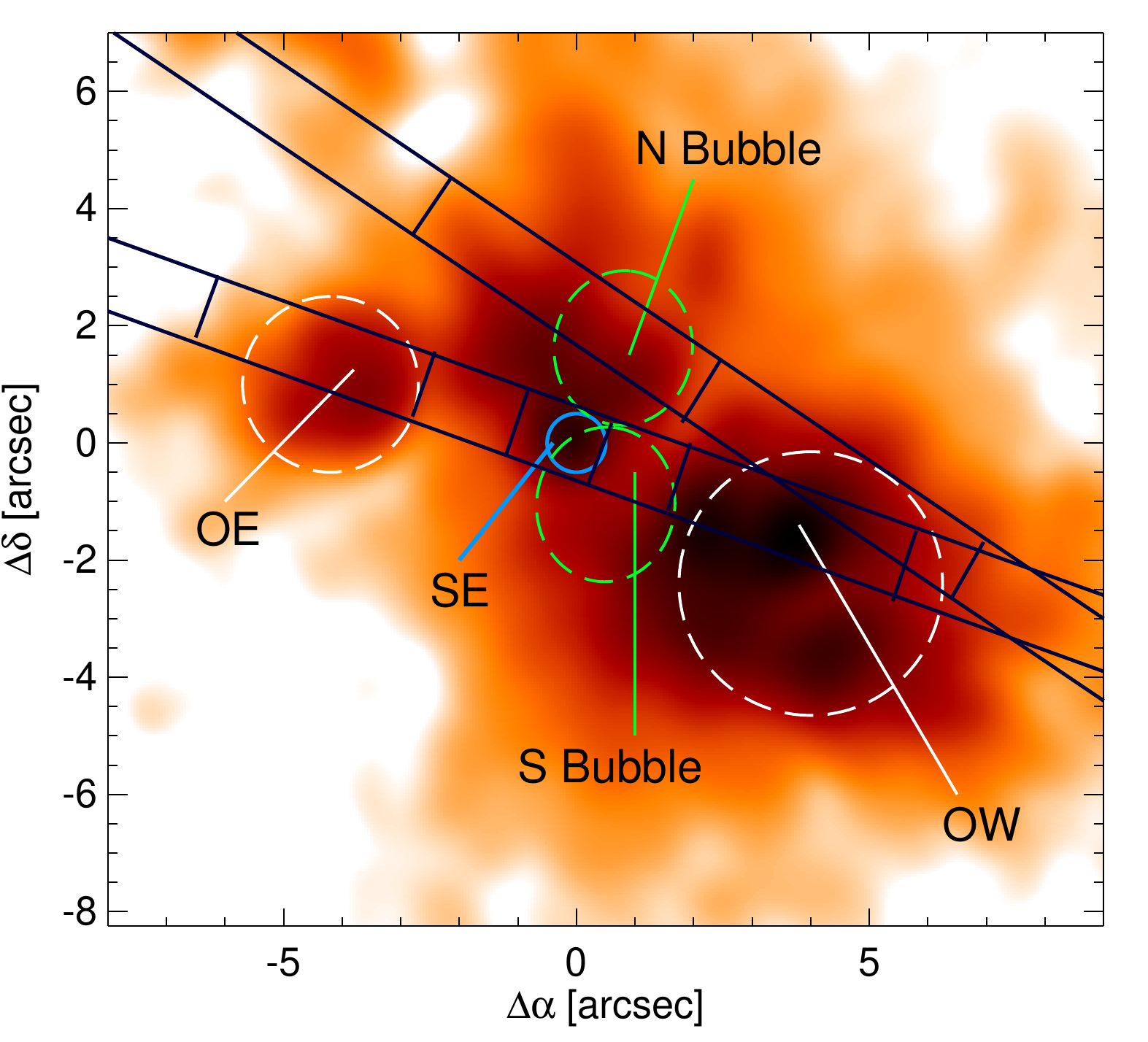}
\caption{Upper panel: {\it HST} continuum subtracted [OIII] image of the galaxy
  showing the location of the slits and the extraction apertures. The images
  have been convolved with a Gaussian of FWHM = 1'' to simulate the seeing
  conditions during the spectroscopic observations. Medium and lower panels:
  these two figures are a zoom into the nuclear region of the galaxy. The
  figures show in detail the main regions of interest that will be later
  analyzed in the paper (see the text for details) and the apertures that cover
  those regions. (A color version of this Figure is available in the online
  journal).}
\label{slits}
\end{figure}

\begin{table}
\centering
\begin{tabular}{llllll}
\hline\hline
Slit Position & Slit PA & Exp time & Airmass & Seeing\\
& degrees& (s) & & (arcsec)\\
\hline
Pos1 & 70 & 6000 & 1.24 & 0.9-1.1\\
Pos2 & 56 & 7200 & 1.14 & 0.9-1.2\\
Pos3$^{\star}$ &  70 & 7200 & 1.24 & 0.9-1.2\\
Pos4$^{\star}$ &  56 & 6000 & 1.14 & 0.9-1.1\\
\hline
\end{tabular}
\caption{Log of the {\it INT} IDS spectroscopic observations of Mrk273 (z =
  0.0373) used for this paper. The seeing (FWHM) was estimated using field stars
  in the acquisition images. \newline $^{\star}$ These two slits (Pos3 and Pos4)
  were placed at exactly the same PA as Pos1 and Pos2 respectively, but were
  shifted by 1 arcsec to the North. For clarity, the slit positions 3 and 4 are
  not shown in Figure \ref{slits}.}
\label{Spec-table}
\end{table}

\subsection{Spectroscopic data}

Long-slit optical spectroscopic observations were taken for Mrk273 in May 2013
using the Intermediate Dispersion Spectrograph (IDS) on the {\it INT} . The
observations were obtained using the R400V grating centered at 5250\AA~with the
RED+2 CCD. The instrumental set-up resulted in a spatial scale of 0.44 arcsec
pix$^{-1}$, and a dispersion of 1.5~\AA~pix$^{-1}$. The useful wavelength range
is $\sim$ 4000 -- 7200~\AA. The data were reduced (bias subtracted, flat field
corrected, cleaned of cosmic rays, wavelength calibrated and flux calibrated)
and straightened before extraction of the individual spectra using the standard
packages in {\sc IRAF} and the {\sc STARLINK} packages {\sc FIGARO} and {\sc
  DIPSO}. The wavelength calibration accuracy, as determined using the mean
shift between the measured and published \citep{Osterbrock96} wavelength of
night-sky emission lines, is $\sim$0.35~\AA~(20 km s$^{-1}$). The spectral
resolution, calculated using the widths of the night-sky emission lines (FWHM),
is 3.1$\pm$0.1\AA~at 6300 \AA ($\sim$145$\pm$5 km s$^{-1}$). The estimated
uncertainty for the relative flux calibration is $\pm$5\%, based on comparison
of the response curves of various spectrophotometric standard stars observed
during the observing runs. Finally, the seeing during the observations, measured
using stars in the acquisition image, was 0.9 -- 1.2 arcsec. Details of the {\it
  INT} IDS observations for Mrk273 can be found in Table \ref{Spec-table}.

For these observations we used 4 slit positions selected to include the main
ionized gas features observed in our ACS-\oxythr~image of the galaxy. Two slits
were placed at PA 70, plus another two at PA 56. The relative shift between each
pair of slits at the same PAs was 1 arcsec to the north. The slits cover the
bright [OIII] emission in the nuclear region of the galaxy as well as a
significant fraction of the faint, very extended [OIII] emission observed
towards the east of the nuclear region.

The extraction apertures for each slit position were selected based on spatial
cuts of the 2D-frames in wavelength ranges chosen to include the
[OIII]$\lambda$5007 and the H$\alpha$+[NII] emission lines. The slit position
and the corresponding extraction apertures for Pos 1 and 2 are shown in Figure
\ref{slits}, overplotted onto the continuum subtracted [OIII] image of the
galaxy. To simulate the seeing conditions during the spectroscopic observations,
the [OIII] image in the figure has been convolved with a Gaussian of FWHM = 1
arcsec.

Since we aim to compare our results with those of previous studies of the
galaxy, it is crucial to know the precise location of the slits. To that aim, we
followed again the same procedure as described in section 2.1.1 to find the
location of the slits on the {\it HST} images with a precision of 5 pixels (0.25
acsec). In addition, we checked the consistency of the flux calibrations for the
entire dataset used in this paper by comparing the {\it INT}-IDS observations
with the {\it HST} images. If we define F$_{\rm IDS}$ as the integrated flux
over a espatial slice extracted from the spectra using wavelength ranges
selected to match those of the {\it HST} filters, we find that F$_{\rm
  HST}$/F$_{\rm IDS}$ is equal to 0.87 and 0.78 for the continuum and [OIII]
observations respectively. Overall, the flux calibrations for the different
datasets used for this paper show a high degree of consistency.

\section{Results}

\subsection{Imaging}

\subsubsection{{\it HST} imaging: the detailed morphology of the ionized gas emission}

The most remarkable feature visible in our emission-line-free, F550M continuum
image of the galaxy (upper-left panel in Figure \ref{HST_frames}) is the well
known extended tail to the south of the galaxy. Note that the F550M continuum
image of the galaxy does not cover the full extent of the tail \citep[$\sim$36
  kpc or $\sim$46''. See for example,][ their Figure 1]{Kim02}. The image also
shows some weak, but significant continuum emission towards the north-east of
the nuclear region and to the west of the tidal tail. The lower-left panel in
our Figure \ref{HST_frames} shows the morphology of the continuum emission in
the central 15$\times$15 arcsec. There is a large number of bright knots, clumps
and other irregular condensations around the brightest ``L-shaped'' structure
close to the N and SE nuclear components. All of these structures are crossed by
dust features, with the most prominent dust lane observed to the east of the
nuclear region, extended from north to south, and then along the tidal tail.

Figure \ref{HST_frames} shows the continuum subtracted \oxythr~image of the
galaxy. This is strikingly different from the continuum image. The first notable
structure in the image is the spectacular system of extended filaments and
clumps to the east of the galaxy. Such clumps and filaments extend $\sim$25''
($\sim$19 kpc) in our \oxythr~image; the average position angle of the
structure, as measured relative to the N nucleus, is $\sim$68$^{\circ}$, and the
range of position angles covered by the structure $\sim$50$^{\circ}$.

To better study the morphology of the [OIII] emission in the nuclear region, the
lower-right panel in Figure \ref{HST_frames} concentrates on the [OIII] emission
from the central 15$\times$15 arcsec of the galaxy. The locations of the N, SW
and SE components are indicated in the figure with open circles. It is notable
that the [OIII] emission is not particularly enhanced at the location of the N
and SW nuclei, likely related to reddening effects. However, coinciding with the
location of the SE component we find the brightest and most compact condensation
of [OIII] emission observed in our images. To estimate the size of this feature
we fitted a 2D Gaussian to the region. The resulting FWHM (FWHM$_{\rm unc}$) was
corrected for instrumental width (I$_{\rm FWHM}$) by subtracting the two values
in quadrature\footnote{The instrumental width (I$_{\rm FWHM}$) was measured
  using the same 2D Gaussian fit for two stars in the field. We obtained I$_{\rm
    FWHM}$ = 0.115$\pm$0.005 arcsec. The FWHM of the SE component before the
  correction is FWHM$_{\rm unc}$ = 0.174$\pm$0.005 arcsec. The corrected FWHM is
  calculated as FWHM$_{\rm corr}$ = $\sqrt[]{FWHM_{\rm unc}^{2} - I_{\rm
      FWHM}^{2}}$.}. The corrected FWHM is FWHM$_{\rm corr}$ = 0.13$\pm$0.02
arcsec (99$\pm$15 pc). Therefore, although the SE component is compact, it is
resolved in our observations.

The lower-right panel in Figure \ref{HST_frames} also shows a schematic of the
nuclear-superbubble model proposed by RV13 in their recent IFS study of the
galaxy. Interestingly, the figure shows a ``U-shaped'' structure of diffuse
[OIII] emission that closely follows the structure of the northern, near side of
the bipolar superbubble (referred here as ``N-Bubble''). However, the southern,
far side of the bubble (``S-Bubble'') is not clearly visible in our image, which
is likely related to reddening effects. Finally, on larger scales, but still
within the central $\sim$15 arcsec, there are two high surface brightness
structures of enhanced [OIII] emission extended $\sim$5 arcsec (4 kpc) to the
east and west of the nuclear region, almost perpendicular (PA$\sim$80$^{\circ}$)
to the RV13 nuclear supperbubble. "Overall, the structure of the [OIII] emission
in the nuclear regions is highly complex. A detailed spectroscopic study of the
different structures described in this section is carried out later in the
paper.

\subsubsection{{\it INT} imaging: low surface brightness emission at large scales}

The {\it HST} images of Mrk273 show no evidence for [OIII] emission at or around
the tidal tail to the south of the galaxy. There are two possible explanations
for this finding. The most straightforward explanation is that there is no
ionized gas at these locations. This would be consistent with the results of
\cite{Rodriguez-Zaurin09}, from their detailed study of the stellar populations
in the galaxy. These authors found that the optical (3000 -- 7500\AA) continuum
of the source along the tidal tail is remarkably uniform, and is adequately
modeled using a stellar population of age $\sim$700 Myr that dominates the
optical emission, with a small contribution from a young (i.e. ionizing) stellar
population.

However, given the size of the resolution element of the ACS-WFC, it is possible
that some low-surface-brightness emission line features are not detected in our
{\it HST} images. In order to detect such faint features with the ACS camera,
one would require substantially longer exposure times than those used for our
observations. An alternative is to use deep, lower resolution images. In this
context, Figure \ref{INT-WFC} shows our deep, H$\alpha$+[NII] image of the
galaxy taken with the Wide Field Camera (WFC) at the {\it INT} telescope. The
seeing during these observations was $\sim$1.6''. Therefore, none of the
detailed nuclear structure visible in the {\it HST} images is distinguishable in
Figure \ref{INT-WFC}. However, the figure shows extended, diffuse emission to
the east of the galaxy extended $\sim$30 arcsec ($\sim$23 kpc) from the nuclear
region and to the west of the southern tail.

Unfortunately we do not have a deep INT/WFC line-free continuum images to allow
accurate quantification of the continuum contribution to the extended structures
visible in the INT narrow-band image. However, we note that the filamentary
structure to the west of the southern tidal tail has a peak surface brightness
that is $\sim$50\% of the that of the tidal tail itself. If this were
predominantly a continuum structure, we would expect it to be visible at a
similar contrast level in our F550M continuum image and HST broad-band images of
the source that are continuum-dominated \citep[e.g. the {\it HST} F435W and
  F814W images presented in][]{Mazzarella93,Hibbard96,Kim02}. The fact that the
diffuse emission to the west of the southern tidal tail is in fact much fainter
in the latter images, provides strong evidence that the western filaments --
that run parallel to the main tidal tail -- are predominantly emission line
structures.

We further note that the western filamentary structure partially follows the
morphology of the extended X-ray nebula reported by \cite{Iwasawa11}. Their
results, based on new {\it Chandra} X-ray observations of the source, are
consistent with the tidal tail being located in front of the soft X-ray nebulae
in our line of sight, such that the soft X-ray emission is partially absorbed by
the cold gas in the tidal tail. Moreover, \cite{Iwasawa11} derived a lower limit
on the column density of $N_{\rm H} \sim$ 10$^{22}$cm$^{-2}$, similar to that of
edge-on galaxy disks, and consistent with other measurements of disk shadowing
\citep{Barber96}. \cite{Iwasawa11} concluded that the source of the soft X-ray
emission is likely to be star formation in the edge-on disk, which would be, a
priori, consistent with our {\it INT}-WFC image. In this scenario, the highly
reddened, young stellar populations in the disk would not make a large
contribution to the visible optical emission at these locations, which would
explain the \cite{Rodriguez-Zaurin09} results for the stellar populations along
the tail. Note however, that while \cite{Iwasawa11} found soft X-ray emission on
both sides of the tidal tail, H$\alpha$+[NII] emission is only detected to the
west of the tail in our {\it INT}-WFC image. Future, deeper H$\alpha$ images of
the galaxy will help to investigate in detail the extended ionized gas emission
around the tail and its relation with the soft X-ray emission at these
locations.

\begin{landscape}
\centering
\begin{table}
\begin{tabular}{llllcccccccccccc}
\hline\hline
Extraction & kin    & FWHM &$\Delta V$ & H$\beta$ & [OIII]$\lambda$5007 &
[OI]$\lambda$6300 & H$\alpha$ & [NII]$\lambda$6583 & [SII]$\lambda$6716,6731\\
Aperture (AP)  & Comp   & \kms & \kms  & erg s$^{-1}$ cm$^{-2}$  &&&&&&\\
(1)    & (2)&  (3) &   (4)  & (5) & (6) & (7) & (8) & (9)& (10) \\   
\hline
IDS-POS1\\
\hline
a$^{\star}$ &N1&158$\pm$26    &-64$\pm$22   &8.36E-16 &13.53   &0.56    &4.17     &5.71      &2.11\\
  &N2&482$\pm$31    &-41$\pm$22  &5.01E-15 &6.41    &0.36    &2.99	  &2.74      &1.52\\
  &B&1458$\pm$157 &-204$\pm$79  &0.33     &1.92    &0.24    &2.295E-15&3.01      &1.18\\
\hline
b &N&212$\pm$30   &-12$\pm$25   &2.99E-16 &11.68     &1.06    &7.73	&7.32      &4.34\\
  &I1&644$\pm$28   &-173$\pm$32  &1.20E-15 &9.00      &0.76    &7.02	&6.54      &3.74\\
  &I2&623$\pm$64   &411$\pm$46   &2.08E-16 &4.36      &1.40    &5.02	&15.07     &2.67\\
\hline
c &N&431$\pm$33   &-16$\pm$21    &2.12E-15 &3.52      &0.79    &3.93	&2.77      &2.64\\
  &B&1407$\pm$59  &111$\pm$33   &5.66E-16 &5.34      &1.94    &6.44	&19.32     &5.72\\
\hline
d &N1&193$\pm$46   &285$\pm$25    &2.87E-16 &2.11      &...    &5.35	&4.31      &1.64\\
  &N2&362$\pm$53   &95$\pm$43    &6.84E-16 &7.35      &0.15    &3.15	&1.41      &0.89\\
  &I&876$\pm$122  &73$\pm$48   &3.70E-16 &8.20      &1.69    &4.09	&8.48      &6.34\\
\hline
e &N&63$\pm$21     &67$\pm$24    &3.43E-16 &7.27      &...     &1.67	&0.65      &...\\
\hline
f$^{\dagger}$ &N&38$\pm$15     &114$\pm$27    & ...&  9.94E-16  & ...  &  ...   &  ...    \\	
\hline
IDS-POS2\\
\hline
g &N1&163$\pm$22    &-13$\pm$23    &9.46E-16 &9.47   &0.89    &5.24   	&6.23      &2.93\\	    
  &N2&392$\pm$43   &-166$\pm$30   &1.78E-15 &5.75   &0.31    &4.80	&3.25      &2.26\\
  &I&856$\pm$63   &-73$\pm$34   &9.53E-16 &12.57  &1.27    &8.36	&16.75     &7.87\\
\hline
h$^{\star}$ &N&149$\pm$23    &-22$\pm$23    &1.26E-15 &0.56   &0.26    &6.44	&1.98      &1.60\\
  &I&500$\pm$26    &94$\pm$23    &5.35E-15 &2.70   &0.87    &6.96	&8.00      &4.48\\
  &B&1518$\pm$116  &-468$\pm$10  &0.33     &0.54   &0.2     &8.57E-15	&1.85     &0.67\\
\hline
i &N&111$\pm$25    &-77$\pm$24    &2.23E-16 &8.97   &...    &3.33	&2.26      &...\\
\hline
j &N&84$\pm$32    &-26$\pm$23    &3.18E-16 &12.14  &...    &4.10	&2.05      &...\\
\hline
k$^{\dagger}$ &N&95$\pm$29     &43 $\pm$27    &...   &1.106E-15 & ...    & ...   & ... & ...\\
\hline
\end{tabular}
\caption {The modeling results for the different kinematic components within the
  extraction apertures. Col (1): aperture label as indicated in Figure
  \ref{slits}. Col (2): The label of the different components as defined in
  Section 3.2.1: N (narrow), I (intermediate) and B (broad). For those apertures
  with two components within the same FWHM range, these are indicated with
  numbers (e.g. N1 and N2 corresponding to the two narrow components detected in
  Ap-a). Col (3) and (4): Instrumentally-corrected line widths (FWHM) and shifts
  ($\Delta V$) measure relative to galaxy rest frame (defined as z = 0.0373,
  RV13). The uncertainties in these values were estimated accounting for the
  uncertainties from the modeling and the systematic calibration errors
  (i.e. the uncertainty in the instrumental width and the wavelength calibration
  error). Column (3): H$\beta$ flux in erg s$^{-1}$ cm$^{-1}$. From Columns
  (4)--(10): ratios of the fluxes of the main emission lines to the H$\beta$
  line. The uncertainties in these ratios are typically $\lsim$15 per cent, as
  estimated accounting for the 5 per cent flux calibration uncertainty, and the
  uncertainties in the model fits themselves. However, there are a some
  particular cases, mainly those kinematic components that make a small
  contribution to the flux, for which the uncertainty can be as high as 50 per
  cent (for example, the B component of the AP-a). These line ratios are not
  corrected for reddening. \newline$^{\star}$ For these two apertures the
  contribution of the broad component (B) to the overall H$\beta$ flux is
  negligible. Therefore, when such component is included in the modeling its
  properties are entirely unconstrained, leading to unrealistic line ratio
  values. For the broad component in these two apertures, the fluxes are given
  relative to the H$\alpha$ line. In addition, when calculating the line ratios
  used later for the diagnostic line ratio diagrams we took the conservative
  approach of assuming an H$\alpha$ to H$\beta$ ratio of 3. Note that, since
  reddening effects are expected to be important at the locations in the galaxy
  covered by these extraction apertures (i.e. the H$\beta$ flux will be in
  reality less than 3 times the H$\alpha$ flux), the [OIII]/H$\beta$ line ratios
  calculated are lower limits of the real values. \newline$^{\dagger}$ For these
  two extraction apertures, only the [OIII]$\lambda\lambda$4959,5007 emission
  lines are detected. The table show the [OIII]$\lambda$5007 flux in units of
  erg s$^{-1}$ cm$^{-1}$. }
\label{model_results}
\end{table}
\end{landscape}

\begin{figure*}
\centering
\includegraphics[width=0.9\textwidth, height=0.75\textheight]{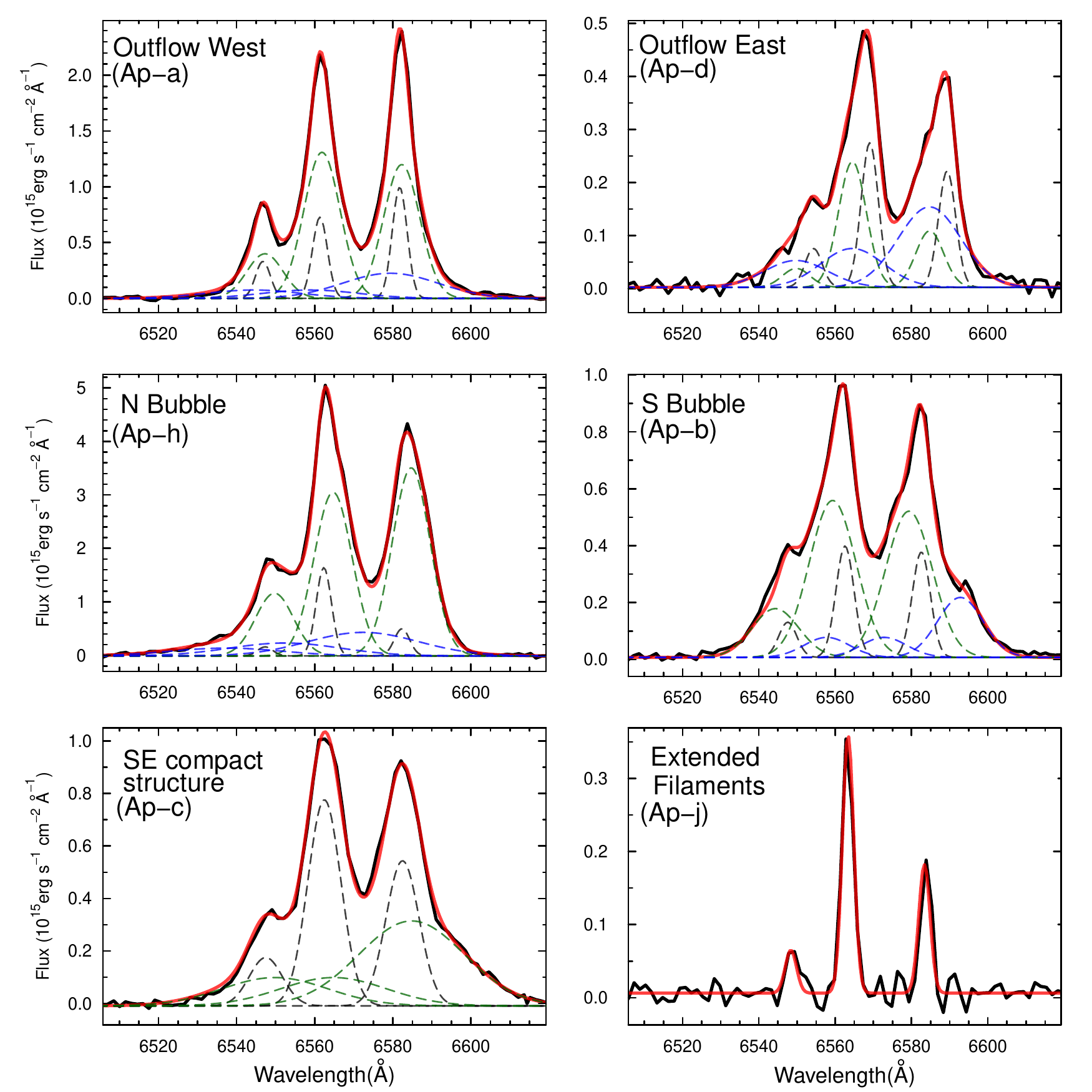}
\caption{The emission line modeling results for the H$\alpha$ ($\lambda_{\rm
    rest}$ = 6563\AA) and the [NII] emission lines ($\lambda_{\rm rest}$ = 6548
  and 6583\AA). The black solid line represents the data, while the red solid
  line represents the overall fit to the emission lines. Black, green and blue
  dashed lines correspond to the different kinematic components in order of
  increasing FWHM. The region of the galaxy corresponding to each of the
  emission line profiles, and the extraction aperture that samples that region,
  are also indicated in the figure. (A color version of this Figure is available
  in the online journal.)}
\label{Ha-profiles}
\end{figure*}

\subsection{Spectroscopy}

\subsubsection{Continuum emission subtraction}

Prior to the modeling of the emission line profiles in the nuclear regions, the
spectra were shifted to the galaxy rest frame ($z = 0.0373$: RV13). Since we aim
to study not only the kinematics but also the ionization mechanisms in Mrk273,
it is important to subtract the underlying stellar continuum prior to modeling
the emission line profiles. This might be an important issue for the H$\beta$
and H$\alpha$ emission lines, which are are the most affected by the stellar
absorption. Unfortunately, our {\it INT}-IDS spectra do not have sufficient
wavelength range coverage to carry out a detailed analysis of the stellar
populations. For this reason, we used the results of spectral synthesis model
fits to the wider-coverage (3000--7500\AA) {\it WHT}-ISIS spectra of this
source, as presented in \cite{Rodriguez-Zaurin09}, to guide our fits to the
stellar continuum of the INT/IDS data.

\cite{Rodriguez-Zaurin09} found that the stellar populations across the full
extension of the galaxy covered by their slit (see Figure \ref{slits}) were
remarkably uniform in terms of their ages, with no evidence for a significant
age gradient. The best fitting models in the central regions (their apertures D,
E and F) comprise a $\sim$0.7--1.0 Gyr ($E(B-V) \leq$ 0.4) stellar population
plus a significant contribution from a reddened $\lsim$50 Myr ($E(B-V) \leq$
1.0) stellar population (see \cite{Rodriguez-Zaurin09} for details). Therefore,
we took the approach of assuming a similar mix of stellar populations to correct
for the stellar continuum the {\it INT}-IDS spectra. This approach was used only
for those apertures that sample regions with significant continuum emission, as
observed in the ACS continuum image of the source. These are: AP-a, b, c, d, g
and h.

The stellar population within the apertures with significant stellar continuum
were modeled using a similar technique to that described in
\cite{Rodriguez-Zaurin09}, but using a smaller number of discrete values for the
age and reddening of the stellar populations (see \cite{Rodriguez-Zaurin09} for
details). Although it is beyond the scope of this paper to perform a detailed
analysis of the stellar populations in the galaxy, we emphasize that models
including a 0.7 Gyr ($E(B-V) = 0.2$) plus a $<$10 Myr ($0.5 < E(B-V) \leq$ 1.0)
stellar population of varying contribution (25--90\%), accurately reproduce the
stellar continumm emission from the galaxy for these 6 apertures.

Overall, we find that subtracting the stellar continuum is important to better
constrain the properties of the different kinematic components, especially for
the H$\beta$ emission line. For example, 3 kinematic components are required to
adequately model the H$\alpha$ or the [OIII]\lala4959,5007 emission lines at
almost all locations in the nuclear regions. However, before subtracting the
stellar continuum, there are cases where only two of these components are
required when modeling H$\beta$. In these cases we often find that the third
kinematic component, typically the one that makes the smallest contribution to
the overall H$\alpha$ emission, clearly emerges once the stellar emission has
been subtracted from the spectra. The change in flux for the kinematic
components detected before subtracting the stellar continuum is small for
H$\beta$, a factor of $\lsim$1.3. In the case of the H$\alpha$ emission line,
the effects of correcting from stellar absorption are negligible.

\subsubsection{Fitting the emission line profiles}

Following continuum subtraction, we used a combination of $DIPSO$ and the IDL
MPFIT \citep{Markwardt09} code to fit Gaussian profiles to the emission
lines. Our modeling approach is described in detail in
\cite{Rodriguez-Zaurin13}. To summarize, we first selected a prominent emission
line that is sufficiently bright and unblended to allow the generation of
a model that could be used to fit all the other emission lines. In this context
the strong [OIII]$\lambda$$\lambda$4959,5007 lines usually have high S/N in our
spectra, are in a region clean from any atmospheric absorption, and are not
blended with other emission lines. We fit these emission lines with the
minimum number of Gaussian components required to produce an acceptable fit. The
number of Gaussian components required in the model for each line was determined
using both an assessment of the reduced Chi squared ($\chi{^2}_{red}$) values
and visual inspection of fits overlaid on the data. For example, models
comprising two kinematic components resulted in $\chi{^2}_{red}$ values of 2.4-4
for Ap-a, b, d and h (See Fig \ref{slits} and \ref{Ha-profiles} ). Including a
third, more kinematically disturbed, component resulted in a significant
decrease in the $\chi{^2}_{red}$, down to values of one or less. A similar
reduction in $\chi{^2}_{red}$ was not seen when we added a second or third
kinematic component in the cases of the SE compact structure (Ap-c), and in
regions covering the extended arc structures to the east (e.g. Ap-j).

In addition, when fitting the [OIII] lines we used three constraints in
accordance with atomic physics: i) the flux ratio between [OIII]$\lambda$5007
and [OIII]$\lambda$4959 was set at 2.99:1 (based on the transition
probabilities); ii) the widths of the corresponding kinematic components of each
line were forced to be equal; iii) the shifts between the corresponding
components of each line were fixed to be 48.0\AA. We will refer to the model
fitted to the [OIII] lines in this way as the ``[OIII] model'' in what follows.

Once the [OIII] model has been derived, we attempted to model the other
prominent emission lines in the spectra with the same kinematic model (velocity
widths and shifts) as {[OIII]}, leaving the relative fluxes in the kinematic
sub-components to vary.  As well as the constraints derived from the {[OIII]}
model, we further constrained the fits to other doublets in accordance with
atomic physics (see RZ13 for details). After a series of iterations, the initial
[OIII] model was refined so that the final kinematic model is the best fitting
model that adequately reproduces all the prominent emission lines in the
spectra. The velocity widths derived from the fits were quadratically corrected
for the instrumental profile, and all linewidths and radial velocity shifts were
corrected to the rest frame of the object (z = 0.0373).

Table \ref{model_results} shows the velocity shifts, widths, line fluxes and
emission line rations, for slit positions Pos1 and Pos2. Note that the line
ratios in the table have not been corrected for reddening. This is because
estimating the reddening is challenging based just on our existing
data. H$\alpha$ is in a blend, and there are potentially degeneracies involved
in the fits that may affect the accuracy of the line ratio measurements,
particularly in the case of the highly complex emission line kinematics present
in the near-nuclear regions of Mrk273.

As mentioned in Section 2.2, two additional long-slit spectra were taken at the
exact same PAs as Pos1 and Pos2 but shifted one arcsec to the North. These are
referred as Pos3 and Pos4 in Table \ref{Spec-table} . For each of these two
additional slit positions we extracted a set of 4 extraction apertures
respectively that sample similar regions to those sampled by Pos1 and Pos2 (but
1 arcsec immediately to the north). The modeling results are, in general,
consistent with those for Pos1 and Pos2 and therefore are not shown in Table
\ref{model_results}.

\begin{figure*}
\centering
\begin{tabular}{cc}
\includegraphics[width=0.4\textwidth]{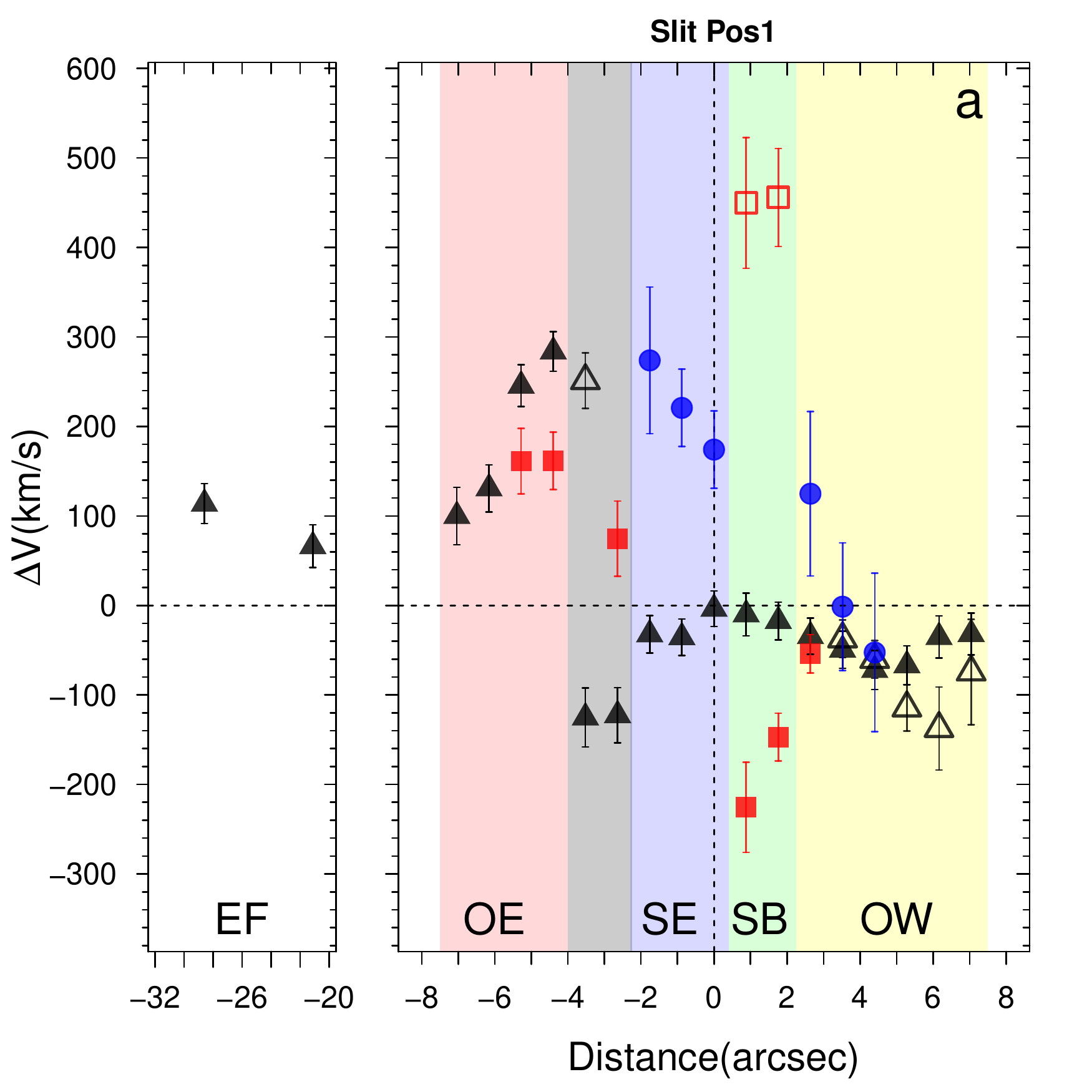}&
\includegraphics[width=0.4\textwidth]{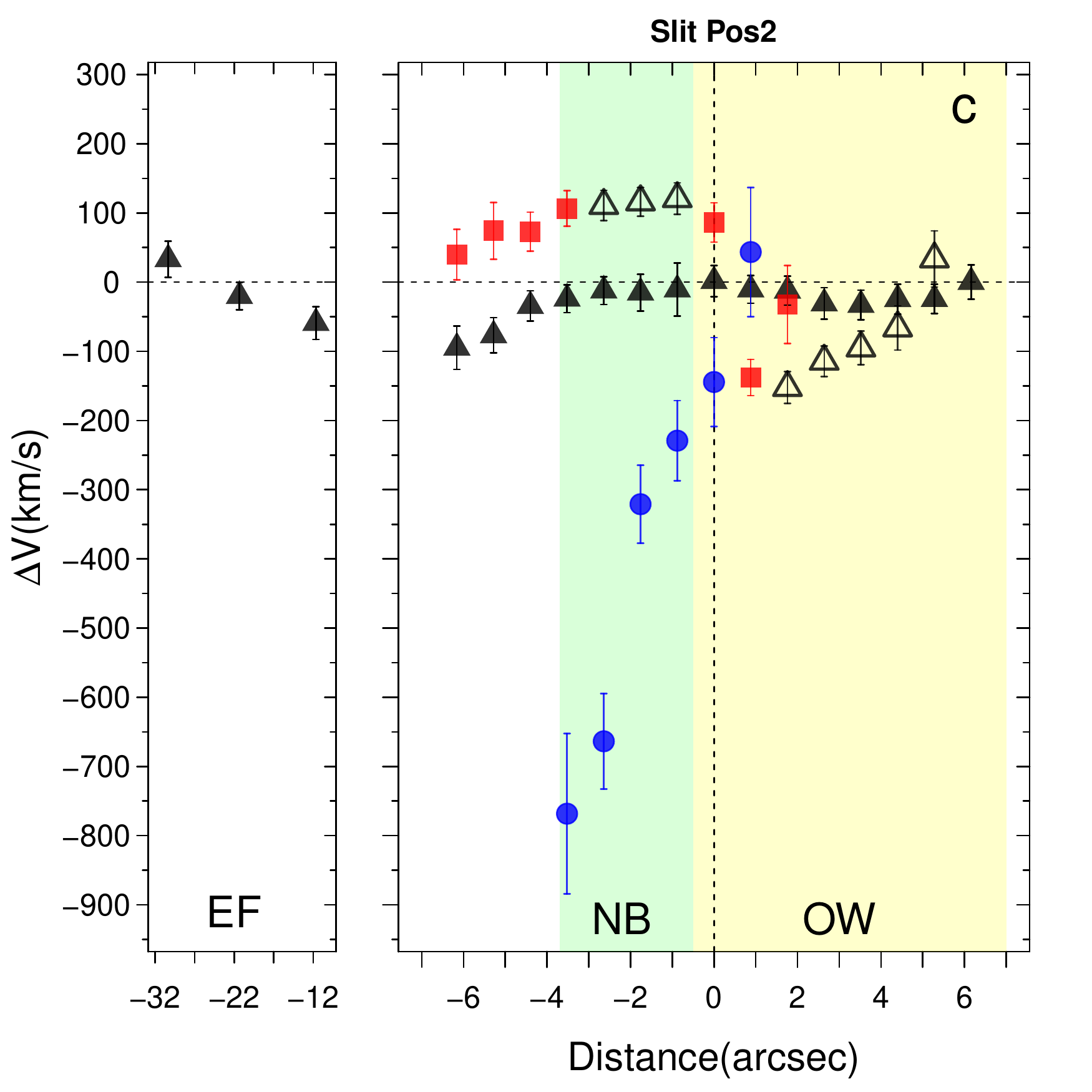}\\
\includegraphics[width=0.4\textwidth]{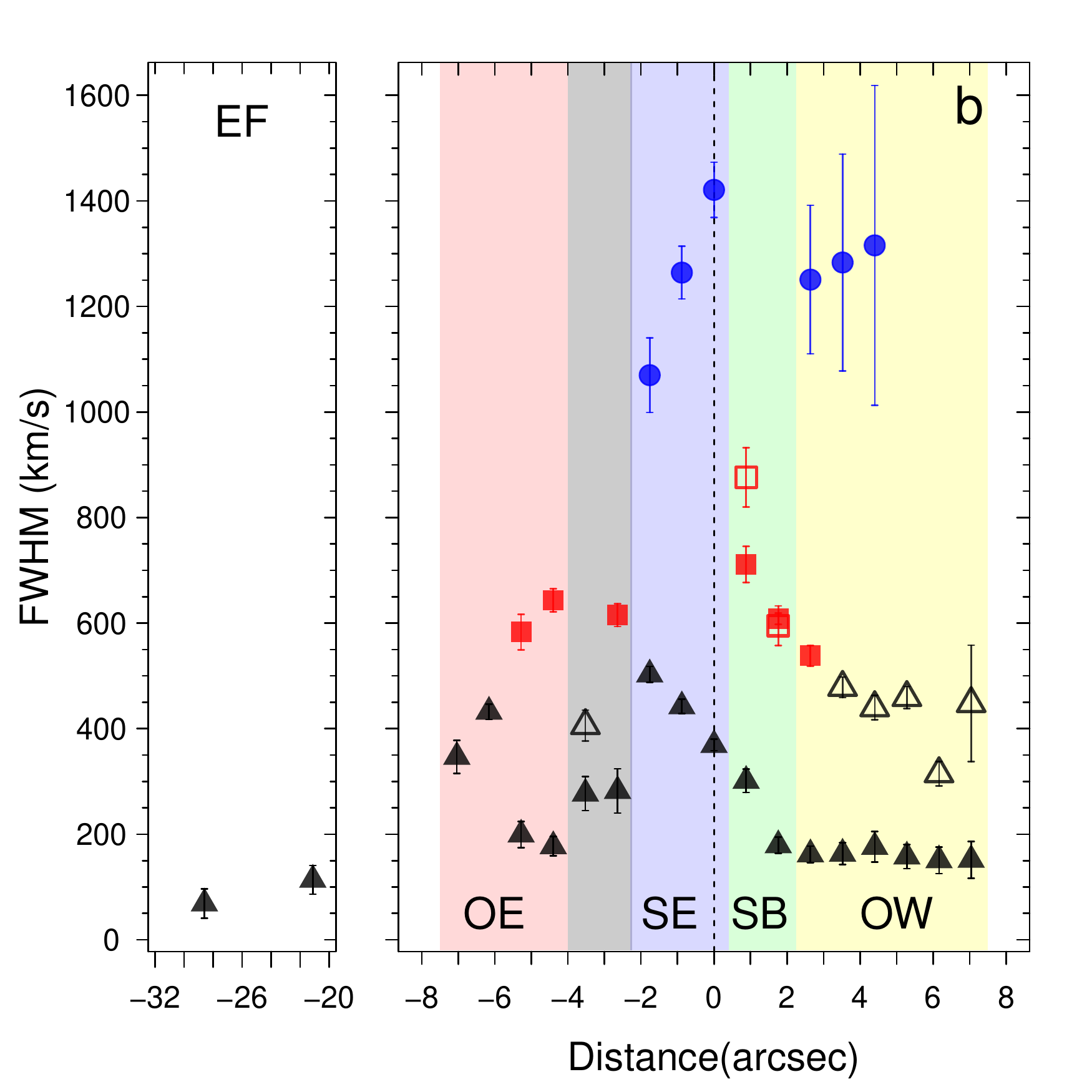}&
\includegraphics[width=0.4\textwidth]{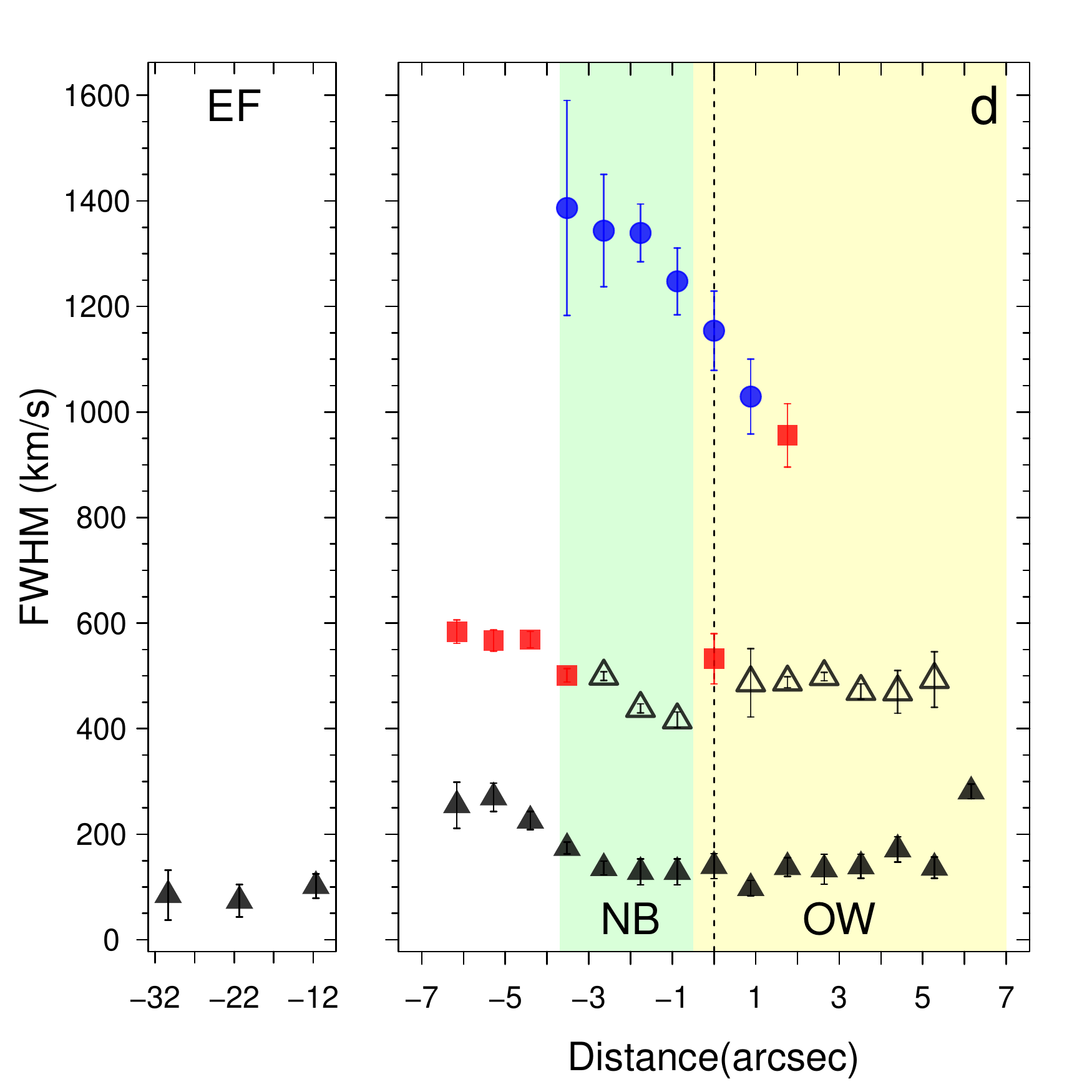}\\
\end{tabular}
\caption{Velocity shifts ($\Delta$V) and widths (FWHM, corrected for
  instrumental width) plotted against the distance from a reference point in the
  galaxy, for Slit Pos1 and Pos2 (see the text for details on the selection D =
  0 arcsec reference point). Extraction apertures of 2 pixels (i.e 0.88 arcsec)
  were used for the Figure, except at the very extended regions, where we used
  the apertures Ap-e and Ap-f for Pos1 and Ap-i, Ap-j and Ap-k for Pos2. Black
  triangles, red squares and blue circles correspond to the narrow, intermediate
  and broad components. To help the reader associate the points in the upper
  panels to those in the lower panels, when more than one kinematic component
  within the same FWHM range is present at a certain location, the second or
  third component (typically the broader) is indicated with an open symbol. For
  example, in Figures \ref{kin-D}a and \ref{kin-D}b, the ionized gas emission at
  D$\sim$7 arcsec is modeled using two kinematic components of widths FWHM =
  152$\pm$25 and FWHM = 456$\pm$106, i.e. both within the narrow FWHM range. To
  be able to unambiguously associate the points in Figure \ref{kin-D}a to those
  in Figure \ref{kin-D}b at this location, the broader component is plotted with
  an open triangle. Finally, the shaded regions indicates the location in the
  galaxy. For Pos1, from west to east: yellow, green, blue and red correspond to
  Outflow-West (OW), S-Bubble (SB), SE compact structure and Outflow-East (OE)
  respectively. In addition, the grey-shaded region corresponds to the prominent
  dust lane observed to the east of the nuclear region. For Pos2, yellow and
  green correspond to Outflow-West and N-Bubble (NB) respectively. The region
  corresponding to the extended filaments (EF) is also indicated in the
  Figure. A color version of this Figure is available in the online journal.}
\label{kin-D}
\end{figure*}

In order to compare with the recent RV13 optical IFS study, which involves fits
to the H$\alpha$+[NII] complex, Figure \ref{Ha-profiles} shows the fits to the
H$\alpha$+[NII] emission lines profile for 6 extraction apertures selected to
sample the main regions of interest, as observed in our {\it HST} images. These
are: Ap-a, Ap-b, Ap-c, Ap-d, Ap-h and Ap-j (see Figure \ref{slits} for their
location in the galaxy). In Figure \ref{Ha-profiles}, the apertures are also
labelled according to the region that they cover. Ap-a and Ap-d sample the
outflowing gas to the west and east of the nuclear region, and therefore are
labelled as ``Outflow-West'' and ``Outflow-East'' respectively. Ap-b and Ap-h
sample the lobes of the nuclear superbubble found by RV13 to the south and north
of the nuclear region and are labelled as ``N-Bubble'' and ``S-Bubble''
respectively. Finally, Ap-c and Ap-j sample the compact SE structure in the
nuclear region and the filaments of clumps of extended emission, and are
labelled as ``SE compact structure'' and ``extended filaments''. See Figures
\ref{slits} and \ref{Ha-profiles} for details.

For ease of reference in the following sections, we use the following scheme to
label kinematic components, based on line widths (FWHM):

\begin{itemize}
\item narrow: FWHM $<$ 500 \kms ;\\ 
\item intermediate: 500 $\leq$ FWHM $<$ 1000 \kms ; \\
\item broad:  FWHM $\geq$ 1000 \kms.  \\
\end{itemize}

\section{Discussion}

\subsection{Ionized gas Kinematics}

Although the spatial resolution of the long-slit observations is not as good as
that of the HST images, they are useful to gain a general idea of the ionized
gas kinematics at different locations in the galaxy. In this context, Figure
\ref{kin-D} shows the velocity shifts ($\Delta$V) and widths (FWHM) along the
slits Pos1 and Pos2. For this figure, we have used extraction apertures of 2
pixels (i.e 0.88 arcsec) except at the very extended regions, where we used the
apertures Ap-e and Ap-f for Pos1 and Ap-i, Ap-j and Ap-k for Pos2. Since the
kinematics of the galaxy are highly disturbed across the slits, the selection of
a reference point (D = 0 arcsec) is rather arbitrary. In the case of Pos1, the
reference point is the centroid of the SE structure (i.e. RA=0 and Dec=0 in
Figures \ref{HST_frames} and \ref{slits}), while for Pos2, the reference point
coincides with the boundary region between Ap-g and Ap-h in Figure \ref{slits}.

Figure \ref{kin-D} shows that the kinematics in Mrk273 are disturbed at almost
all locations in nuclear regions of the galaxies. For example, the narrow and
intermediate components across Pos2 are well organized around the reference
point (D=0 arcsec). However, a third kinematic component emerges as we approach
that point, and extends into the so called N-Bubble region. Consistent with the
results of RV13, we find extreme velocity shifts and widths of up to
$\Delta$V$\sim$-1000 km s$^{-1}$ and FWHM$\sim$1400 km s$^{-1}$ respectively, at
these locations. In contrast, the emission line profiles along slit Pos1 change
drastically from one region to another. The kinematic properties of the
narrowest component (black-filled triangles) are relatively uniform across the
region of $\sim$10 arcsec that expands from Outflow-West to the prominent dust
lane\footnote{Given the significant reddening effects, the kinematic results
  within the dust lane (i.e. the grey-shaded area) should be interpreted with
  caution.}. However, across this region, 1 or 2 additional kinematic components
with substantially different widths and shifts are required to adequately model
the emission line profiles.

Finally, Figure \ref{kin-D} also show the velocity shifts and widths of the
ionized gas in the so-called extended filaments (Ap-e, Ap-f, Ap-i, Ap-j and Ap-k
in Figure \ref{slits}). Interestingly, the emission lines at these locations
show small velocity shifts and have the narrowest FWHM values (38--111 km
s$^{-1}$). In addition, the figure shows the presence of a positive velocity
gradient across the filaments. This is further confirmed using the results from
slit Pos3 and Pos4, which sample a larger fraction of the emission from the
filaments observed in our ACS images. 

Overall, Figure \ref{kin-D} illustrates the complexity of the gas kinematics in
Mrk273.

\begin{figure*}
\begin{tabular}{cc}
\hspace*{-0.4cm}\includegraphics[width=0.45\textwidth]{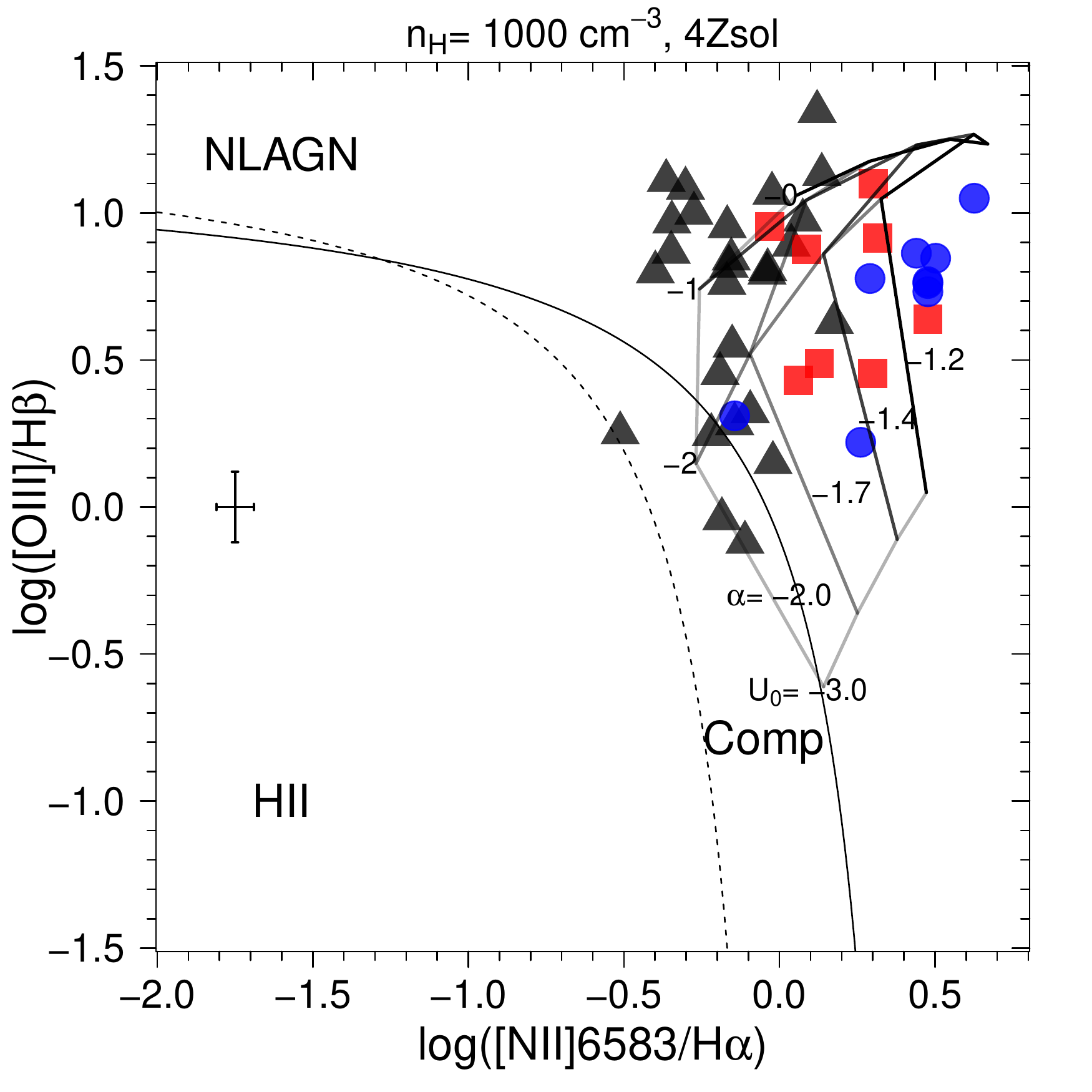}&
\includegraphics[width=0.45\textwidth]{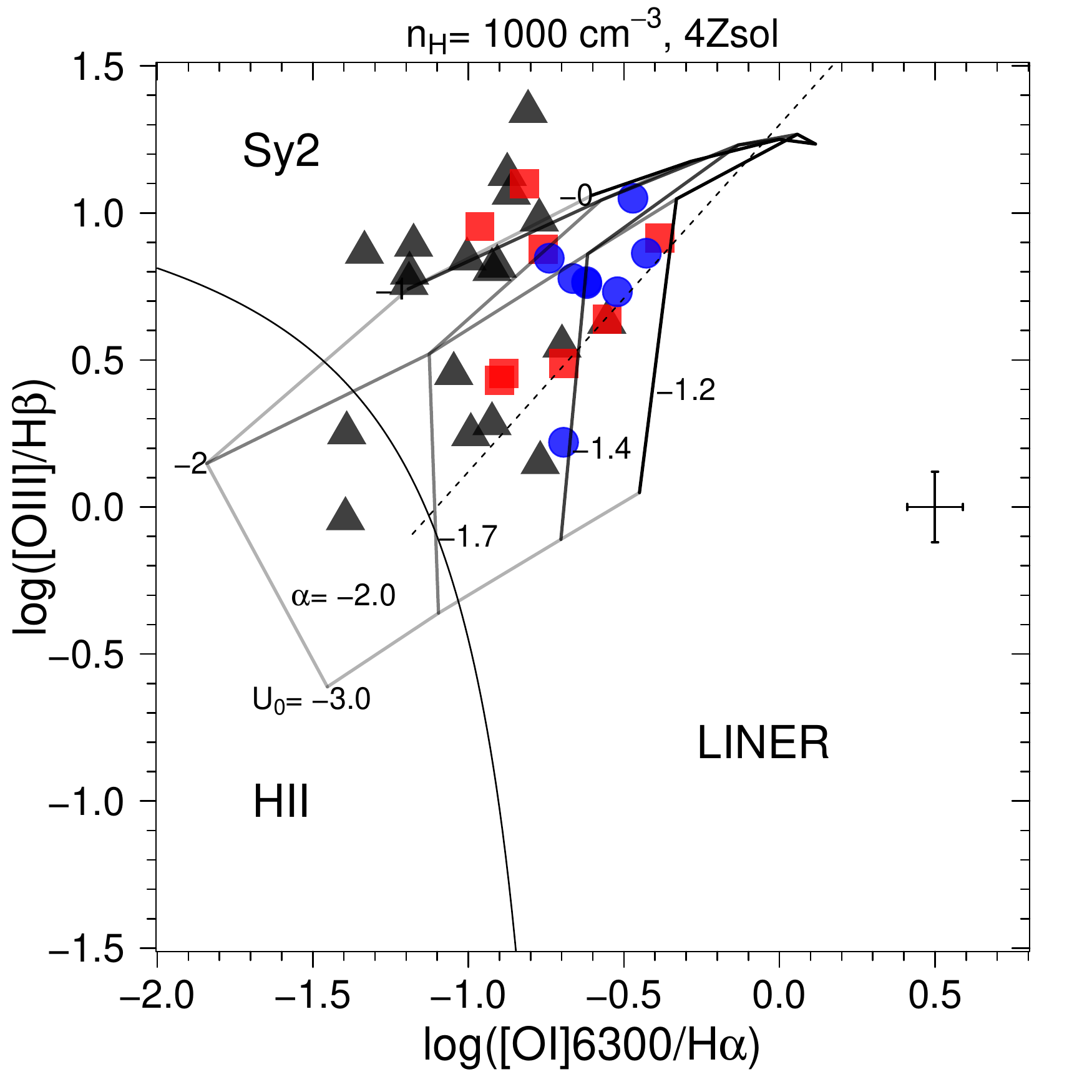}\\
\includegraphics[width=0.45\textwidth]{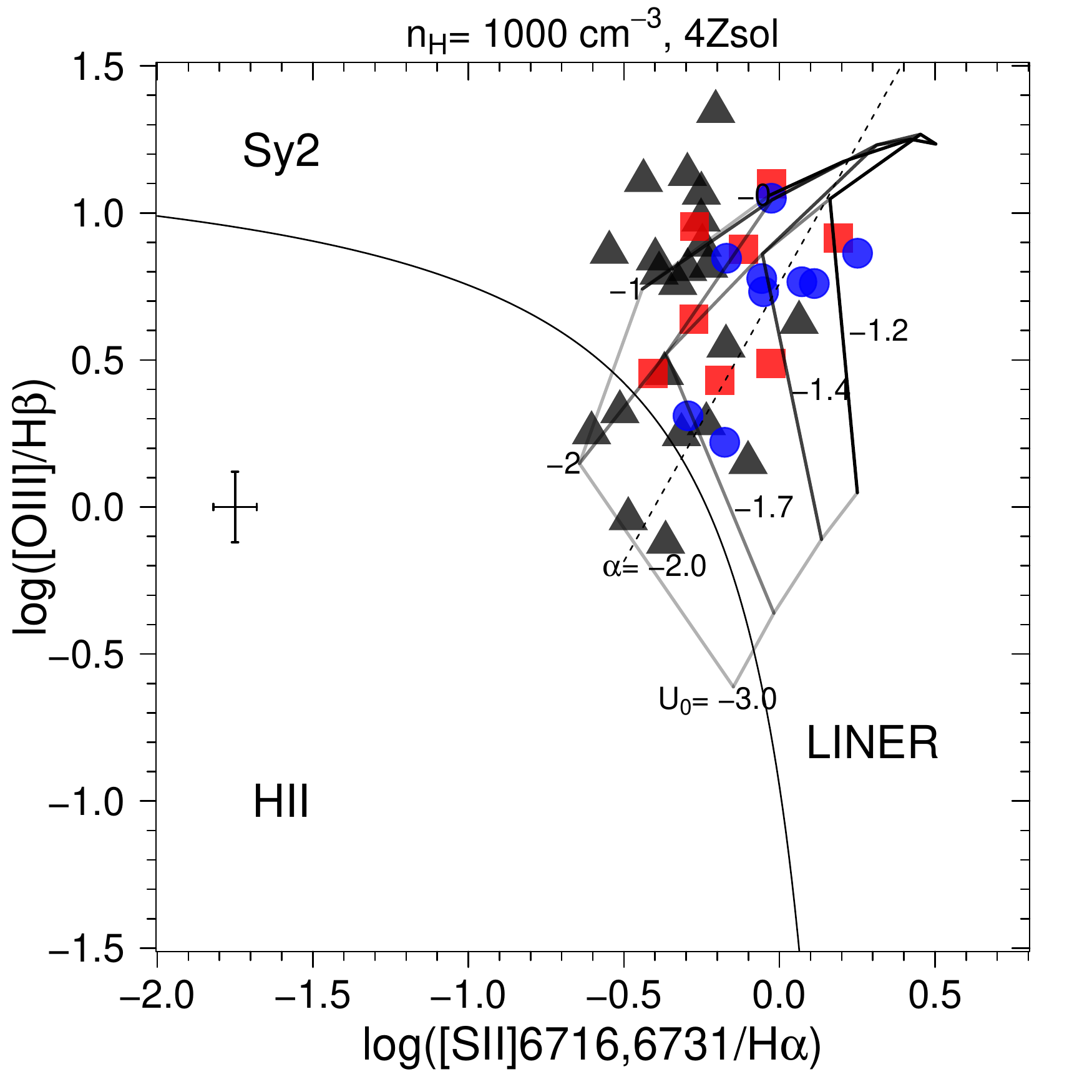}
\end{tabular}
\caption{Standard optical diagnostic diagrams showing the classification scheme
  of \cite{Kewley06}. The solid curves are the theoretical ``maximum starburst
  line'' derived by \cite{Kewley01} as an upper limit for star-forming galaxies,
  and the empirical boundary lines between Seyfert 2 galaxies and LINERs. The
  dashed curve on the [NII] diagram is the \cite{Kauffmann03} semi-empirical
  lower boundary for the star forming galaxies. The \cite{Groves04a} grids of
  dusty, radiation pressure-dominated models are also plotted in the
  Figure. These grids have been generated assuming four times solar metallicity
  (4Z$_{\odot}$) and hydrogen density of n$_H$ = 1000 cm$^{-3}$ (see the text
  for a justification on the selection of these parameters). Gridlines
  corresponding to five values of ionizing parameter ($U_0$ = 0,-1,-2,-3,-4) and
  four values of power law index (F$_{\nu} \propto$ ${\nu}^{\alpha}$, $\alpha$ =
  -1. 2, -1.4, -1.7, -2.0) are shown in the figure. To help the reader follow
  the gridlines these are grey-coded from ``light-grey'' to black, with
  light-grey and black corresponding to the lowest and highest values of $U_0$
  and $\alpha$ respectively. Over-plotted on the diagrams are the results of our
  kinematic study. Black triangles are the line ratios corresponding to the
  narrow components, red squares correspond to the intermediate component and
  blue circles represent the broadest kinematic components. For clarity, no
  individual error bars are shown in the Figure. However, to give an idea of the
  uncertainty associated to the line ratio measurements a cross symbol indicates
  the median of all line ratio errors in each respective diagram.  (A color
  version of this Figure is available in the online journal.) }
\label{BPT-Groves}
\end{figure*}

\subsection{Ionization Mechanisms}

\subsubsection{AGN and starburst photoionization}

\begin{figure*}
\begin{tabular}{cc}
\hspace*{-0.4cm}\includegraphics[width=0.45\textwidth]{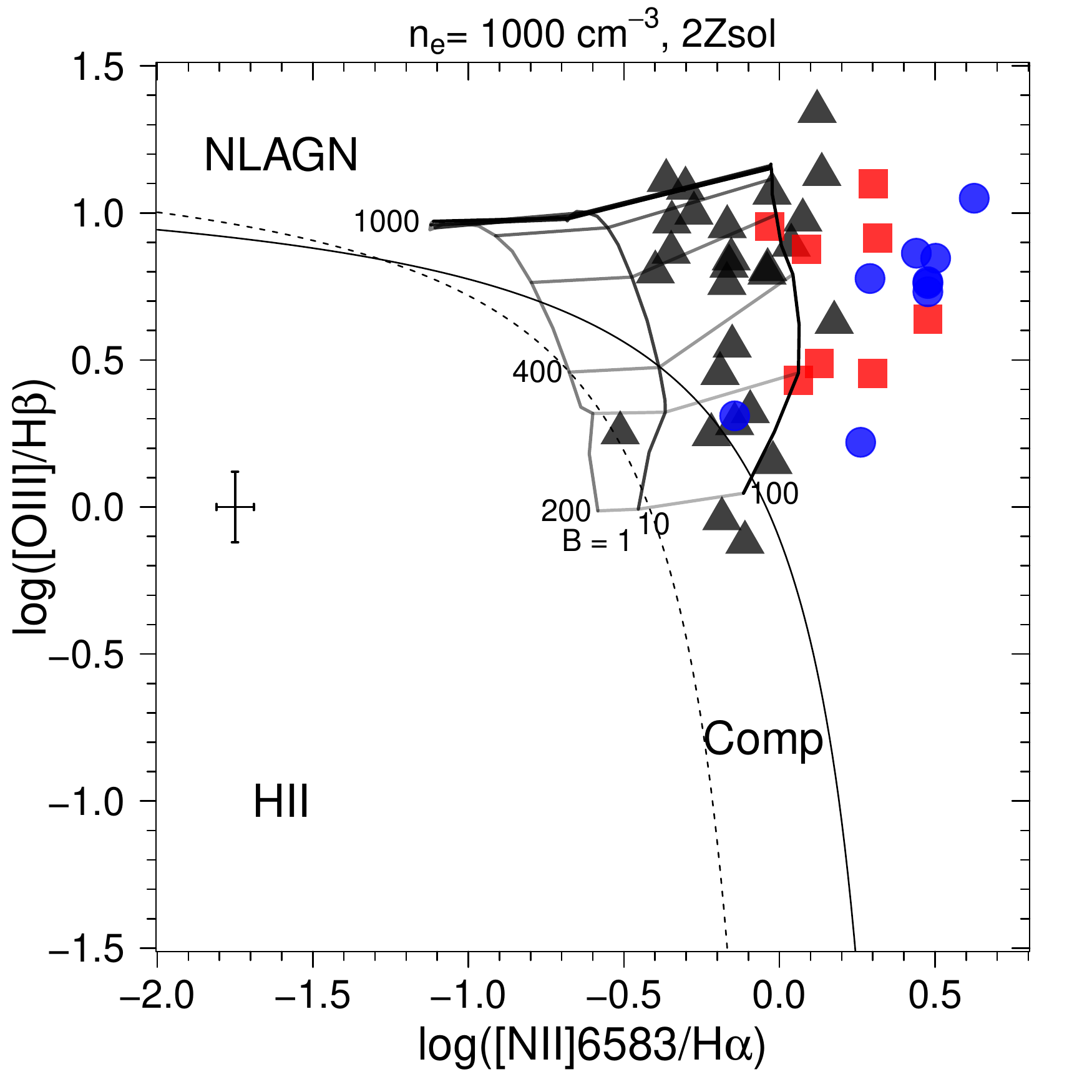}&
\includegraphics[width=0.45\textwidth]{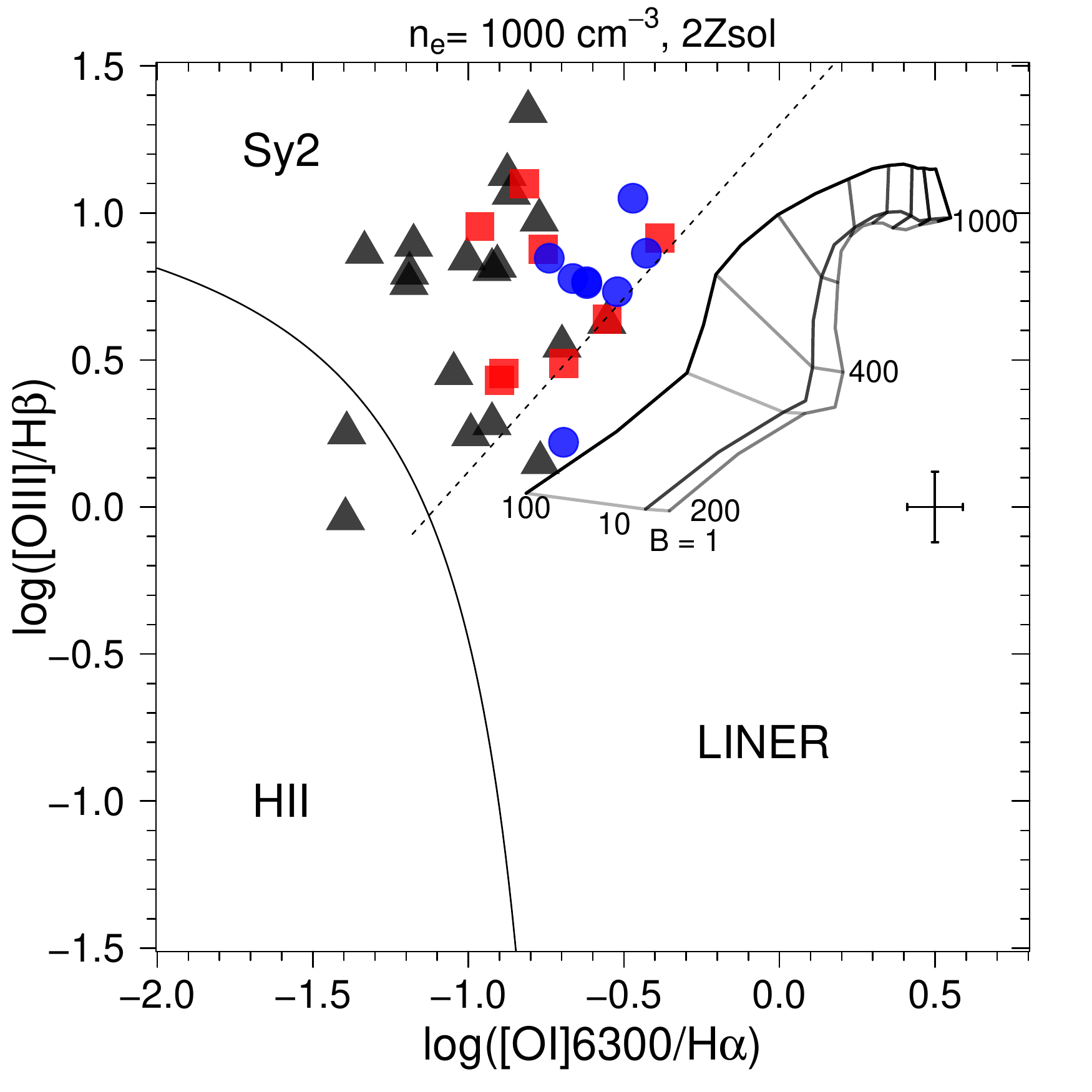}\\
\includegraphics[width=0.45\textwidth]{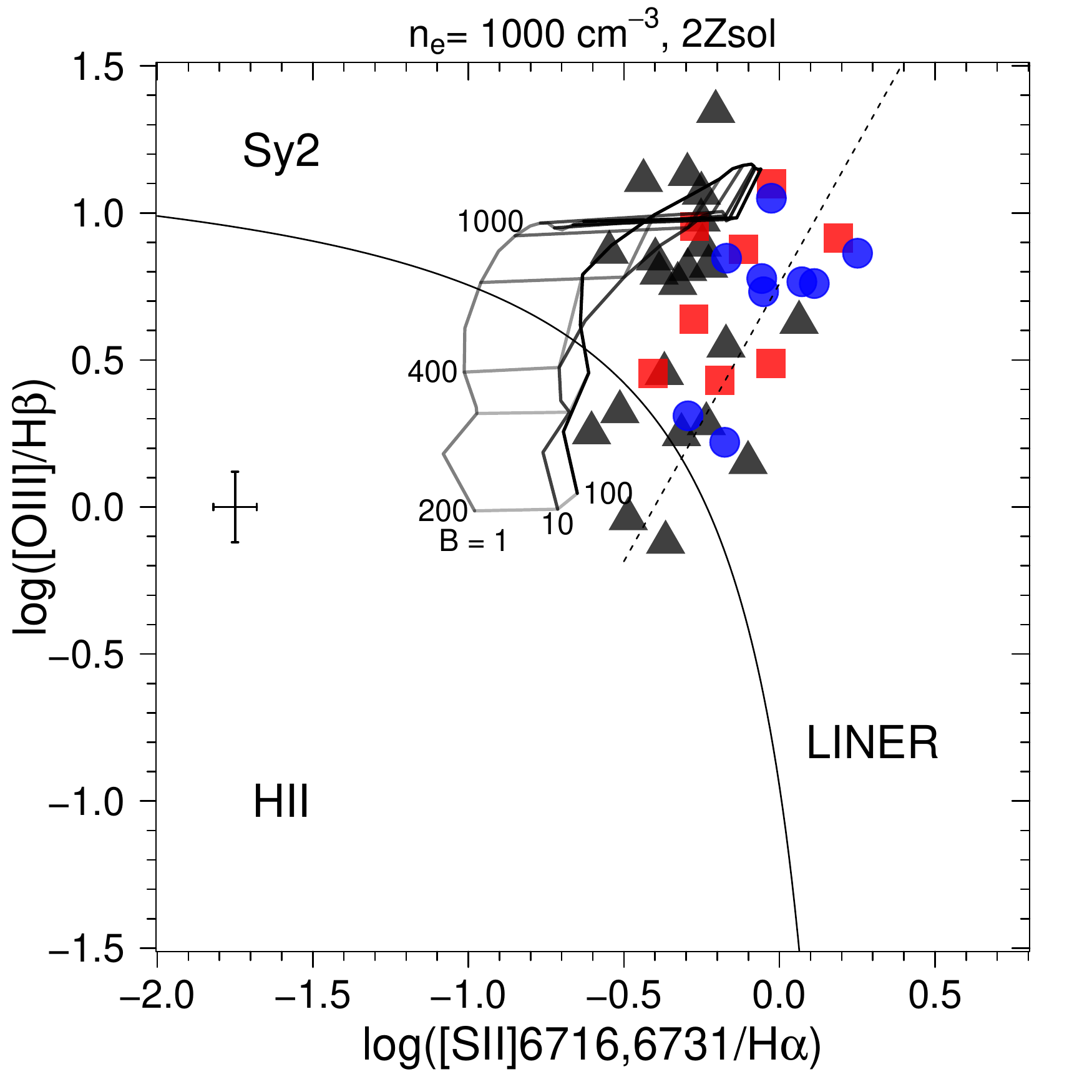}
\end{tabular}
\caption{Same as Figure \ref{BPT-Groves} but showing the \cite{Allen08} grids of
  shock-ionization models. These grids have been generated assuming a
  shock+precursor model with twice solar metallicity and pre-shock density of
  $n_e$=1000 cm$^{-3}$. Gridlines corresponding to 3 values of magnetic field
  ($B$ = 1, 10 and 100 $\mu$G) and 9 shock velocity values ($v_s$ = 200, 300,
  400, 500, 600, 700, 800, 900 and 1000 km s$^{-1}$) are shown in the
  figure. $v_s$ increases always from bottom to top with 3 values (200, 400 and
  1000 km s$^{-1}$) indicated in the figure. As in the case of Figure
  \ref{BPT-Groves}, the gridlines these are grey-coded from ``light-grey'' to
  black, with light-grey and black corresponding to the lowest and highest
  values of $B$ and $v_s$ respectively. The symbols and colors are the same as
  in Figure \ref{BPT-Groves}. (A color version of this Figure is available in
  the online journal.) }
\label{BPT-Allen}
\end{figure*}

In this section we investigate the nature of the ionizing source(s) responsible
for the emission lines observed at the different locations in Mrk273. Figure
\ref{BPT-Groves} shows the diagnostic line ratio diagrams \citep[][hereafter
  BPT/VO87 diagrams]{Baldwin81,Veilleux87} along with the optical classification
scheme of \cite{Kewley06}. Overplotted on the figure are the results from our
kinematic study. For this figure we have used the extraction apertures shown in
Table \ref{model_results} plus the 8 apertures extracted for Slit Pos3 and
Pos4. In addition we also used the \cite{Rodriguez-Zaurin09} {\it WHT}-ISIS
spectrum of the source and modeled the emission lines for their nuclear
extraction apertures (apertures D, E and F in their paper). Overall, a total of
22 apertures, that sample almost the entire nuclear emission, and a large
fraction of the emission at larger scales, were used to generate the plots in
Figure \ref{BPT-Groves}. The black triangles, red squares and blue circles
correspond to the narrow, intermediate and broad components respectively. For
clarity, no individual error bars are shown in the Figure. However, to give an
idea of the uncertainty associated to the line ratio measurements a ``cross
symbol'' indicates the median of all line ratio errors in each respective
diagram. Individual error bars are shown later in the section, when a detailed
analysis of some regions of interest is performed.

In the first place, it is notable that the line ratios measured at almost all
locations in the galaxy are consistent with a Sy2 classification, although a
significant fraction of the points in Figure \ref{BPT-Groves} falls close to the
Sy2/LINER limiting region. We find that the intermediate and broad components
tend to have higher [NII]/H$\alpha$, [OI]/H$\alpha$ and [SII]/H$\alpha$ line
ratios than the narrow components. Interestingly, such trend is not observed for
the [OIII]/H$\beta$.

With the aim of better understanding the nature of ionization mechanisms
responsible for the measured line ratios, Figure \ref{BPT-Groves} also shows the
\cite{Groves04a} dusty, radiation pressure-dominated photoionization models for
NLR in AGN. For these models we assume a hydrogen density n$_H$ = 1000 cm$^{-3}$
and 4 times solar abundance (4Z$_{\odot}$). This latter value was chosen based
on the recent results of RZ13. These authors found that the line ratios derived
for the ULIRGs in their sample were consistent with gas of super-solar
abundances (4Z$_{\odot}$) photoionized primarily by an AGN. A grid of models for
various values of the ionizing parameter ($U_0$ = 0, -1, -2, -3, -4) and
ionizing continuum SED power-law indicies (F$_{\nu} \propto$ ${\nu}^{\alpha}$,
$\alpha$ = -1.2, -1.4, -1.7, -2.0) is shown in the figure.

In general, AGN photoionization models reproduce the emission line ratios well,
in the sense that there is good consistency in the relative positions of the
points and the models between the different diagrams. Interestingly, the biggest
disagreement between the models and the observed line rations occurs for the
narrow kinematic components, for which the degeneracy issues mentioned in
Section 3.2 are least important (i.e. the flux associated with these components
is better constrained). Finally, We have explored models with lower
metallicities (Z$_{\odot}$ and 2Z$_{\odot}$), and higher and lower densities
(100 and 10000 cm$^{-3}$), and the level of agreement not only does not improve,
but it is worse.

\subsubsection{Shock Ionization}

As well as photoionization by AGN and starburst, it is possible that line ratios
plotted in Figure \ref{BPT-Groves} might be explained in terms of ionization by
fast radiative shocks \citep[][]{Dopita95,Dopita96,Groves04a,Allen08}. Indeed
shocks are a plausible mechanism for accelerating the gas to the velocities we
measure in the nuclear regions of Mrk273. The shock models predict a series of
line ratios for a range of magnetic field strengths ($B$), electron densities
(n$_e$), abundances and shock velocities ($v_s$). In addition, the predicted
line ratios depend on the geometry of the shock, i.e. the presence of a
photoionized precursor \citep[see][for a detailed discussion]{Allen08}. In this
context, \citet{Dopita95} and \cite{Allen08} found that the shock+precursor
models produce a better fit to the line ratios measured for their samples of Sy2
galaxies. Furthermore, given the typical velocities associated with the ionized
gas ($v >$ 150 km s$^{-1}$), one would expect the photoionization front to
expand and form a precursor HII region ahead of the shock \citep[see][for a
  detailed discussion]{Dopita95,Allen08}. For this reasons, we decided to use
models with precursor for the work presented in this section.

We note that RZ13 found that, in general, shock models fail to reproduce the
emission line ratios measured in the nuclear regions of the galaxies in their
sample of Sy-ULIRGs, unless some extreme parameter values are used
(e.g. pre-shock density of 1000 cm$^{-3}$). With this in mind, Figure
\ref{BPT-Allen} shows the \cite{Allen08} models with twice solar metallicity
(2Z$_{\odot}$) and a pre-shock density of 1000 cm$^{-3}$.

\begin{figure}
\includegraphics[width=0.475\textwidth]{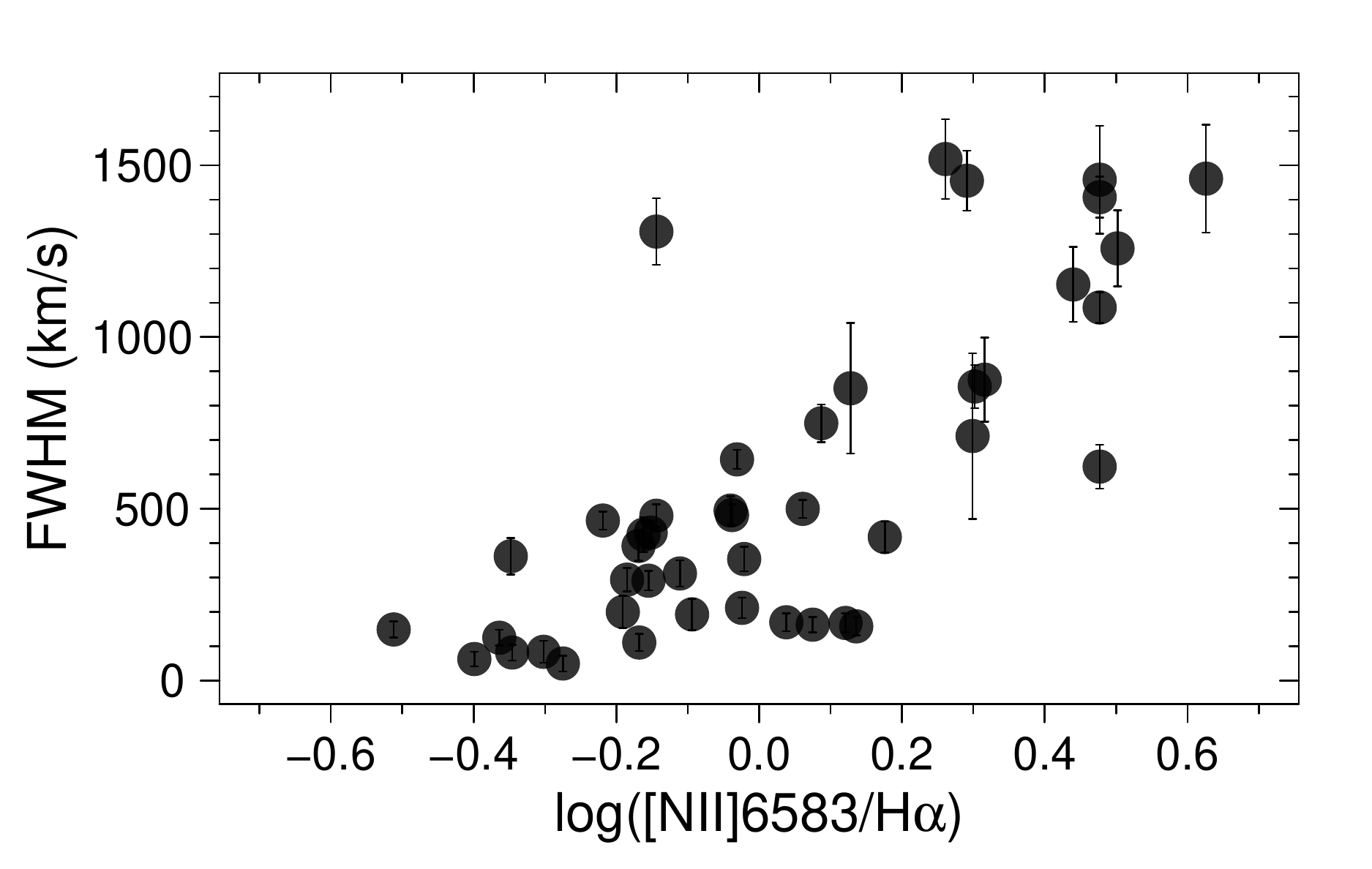}\\
\includegraphics[width=0.475\textwidth]{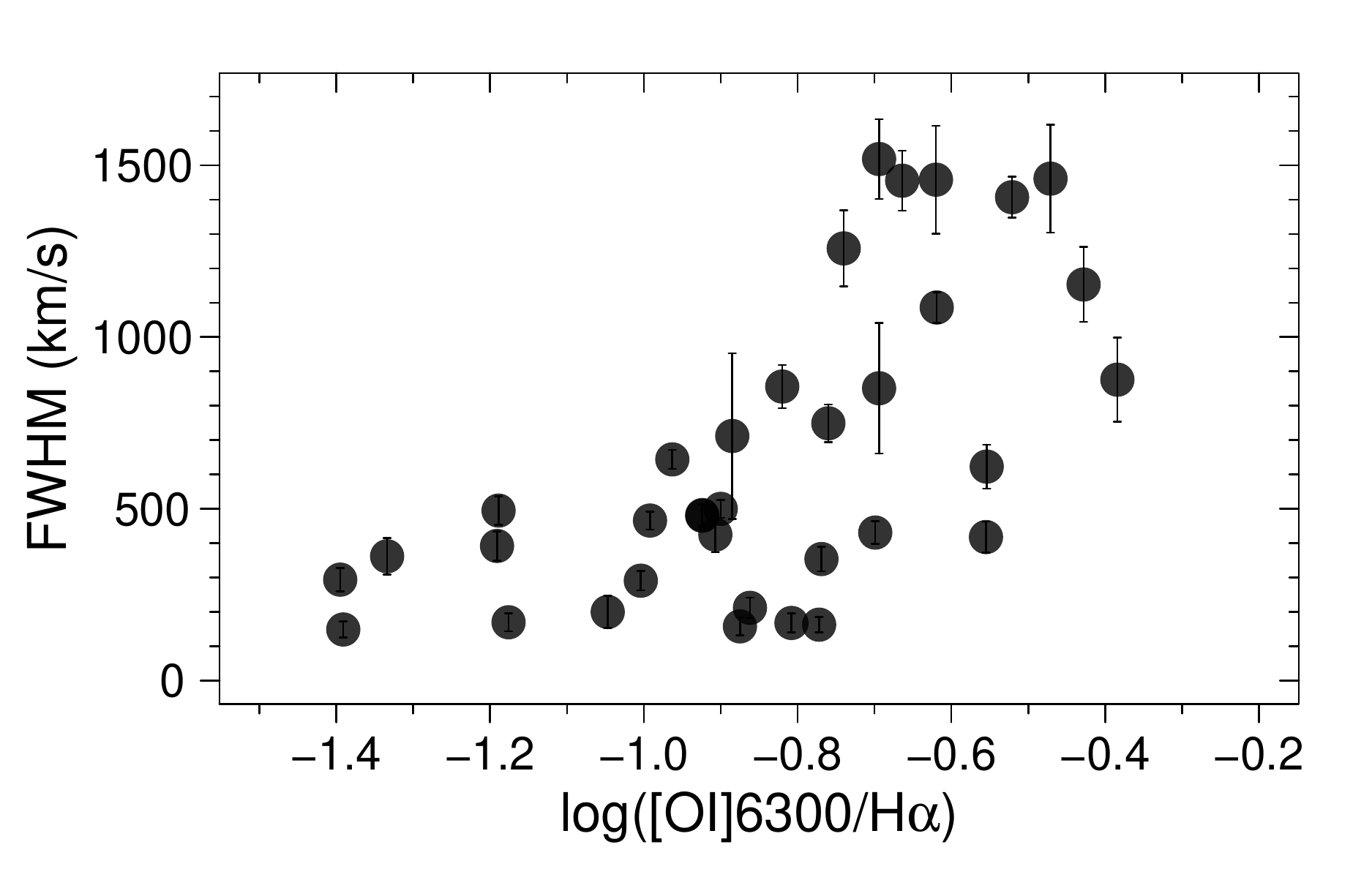}\\
\includegraphics[width=0.475\textwidth]{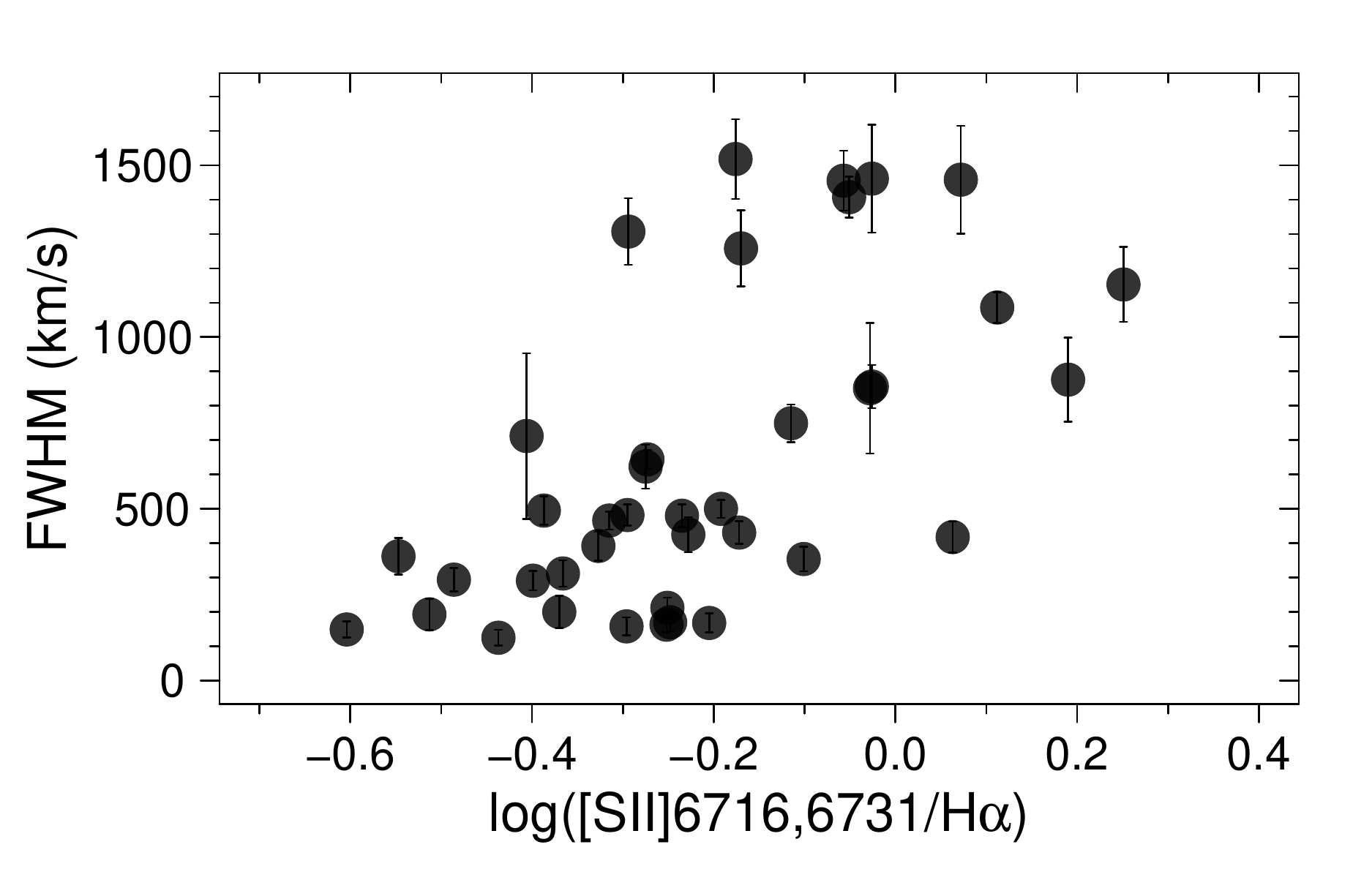}\\
\caption{Velocity widths (FWHM) of the kinematic components at the different
  locations in the galaxy plotted against the line ratio values.}
\label{fwhmvslr}
\end{figure}

The value of the pre-shock density is one of the largest uncertainties in these
models. For example, RZ13 carried out a spectroscopic study of a sample of
Sy-ULIRGs and measured electron densities (n$_{e}$) of $\sim$1000 cm$^{-3}$
(although n$_{e}$ can be substantially higher in some extreme cases: see RZ13
for a discussion on the uncertainties associated with the electron density
measurements). Assuming that these densities are associated with the compressed
post-shock gas in the kinematically disturbed emission line components, then the
pre-shock densities could be as low as 10 -- 100 cm$^{-3}$, since the
compression factor in the cooled, post-shock gas can be high
\citep[$\sim$10-100, see][]{Dopita95}. Note that pre-shock densities of
1000cm$^{-3}$ would lead to post-shock densities of at least 10000
cm$^{-3}$. Therefore, a pre-shock density of 1000cm$^{-3}$ is indeed an extreme
assumption.

Figure \ref{BPT-Allen} shows that, even using the extreme pre-shock density of
1000 cm-3 (as assumed by RZ13), the majority of the measured line ratios do not
fall within the region covered by the grids of shock models in the case of the
[OIII]$\lambda$5007/H$\beta$ vs [NII]$\lambda$6583/H$\alpha$. The disagreement
is even higher for the other two diagrams, where practically none of the
emission lines are covered by the gridlines. As in the previous section, we have
exhaustively explored the parameter space ($B$, $v_s$, $n_e$ and abundance). The
agreement between the models and the measured emission lines improves when
considering the extreme of $B$ = 1000 $\mu$G (this is the highest $B$ value
included in the \cite{Allen08} models). However, such enormous magnetic fields
are unlikely to be found in starburst galaxies \citep{Thompson06}. Similarly,
models with no precursor reproduce better the observed emission line
ratios. However, as we mentioned before, given the high shock velocities
indicated by the emission line kinematics, the presence of precursor HII regions
ahead of the shock is expected.

Finally, it is possible to further test the importance of shock ionization in
Mrk273 by investigating whether there are correlations between the emission line
kinematics and the line ratios. Figure \ref{fwhmvslr} shows the line widths
(FWHM) of the kinematic components at the different locations in the galaxy
plotted against their corresponding line ratios. Shock models predict a strong
correlation between these two quantities. Therefore, these plots have been used
in the past to indicate the presence of shock ionization
\citep[e.g.][]{Armus89,Dopita95,Veilleux95,Monreal-Ibero06,Rich11}. In the case
of Mrk273, we find that the FWHM and line ratios are significantly correlated,
with Spearman's rank correlation coefficient ($\rho$) of 0.69, 0.64, 0.63 for
the [NII]/H$\alpha$, [OI]/H$\alpha$ and [SII]/H$\alpha$ plots respectively, and
p-value $<<$ 0.01 for all three plots (assuming a t-distribution). However, as
indicated from the $\rho$ values, the correlations are not particularly strong.

We further note that shocks may not be a unique explanation of any correlations
between line width and line ratios. For example, it is possible that the
emission lines in the kinematically disturbed regions were initially accelerated
either by shocks or slow entrainment in an AGN-driven wind, but then
photoionized by the AGN. In contrast, the more kinematically quiescent gas (with
low FWHM) has not been accelerated. and is part photoionized by stars or by the
AGN at higher ionization parameter (because it has a lower density due to the
lack of shock acceleration). This combination of components could result in a
correlation between line width and ionization state, even if the more
kinematically disturbed components are currently energized by the AGN
photoinization.

Overall, while the line ratios are generally more consistent with AGN
photoionization than they are with shocks, the correlations that we find between
the line widths and the emission line ratios suggest that shocks may contribute
at some level. Further work, using fainter diagnostic emission lines including,
for example, [OIII]$\lambda$4363 and HeII$\lambda$4686, will be required to
determine the contribution of shocks to the line emission in the nuclear regions
of Mrk273 in a definitive way.

\subsection{Mrk273: understanding the relation between the nuclear and extended
  ionized gas emission}

\begin{figure*}
\begin{tabular}{cc}
\hspace*{-0.4cm}\includegraphics[width=0.45\textwidth]{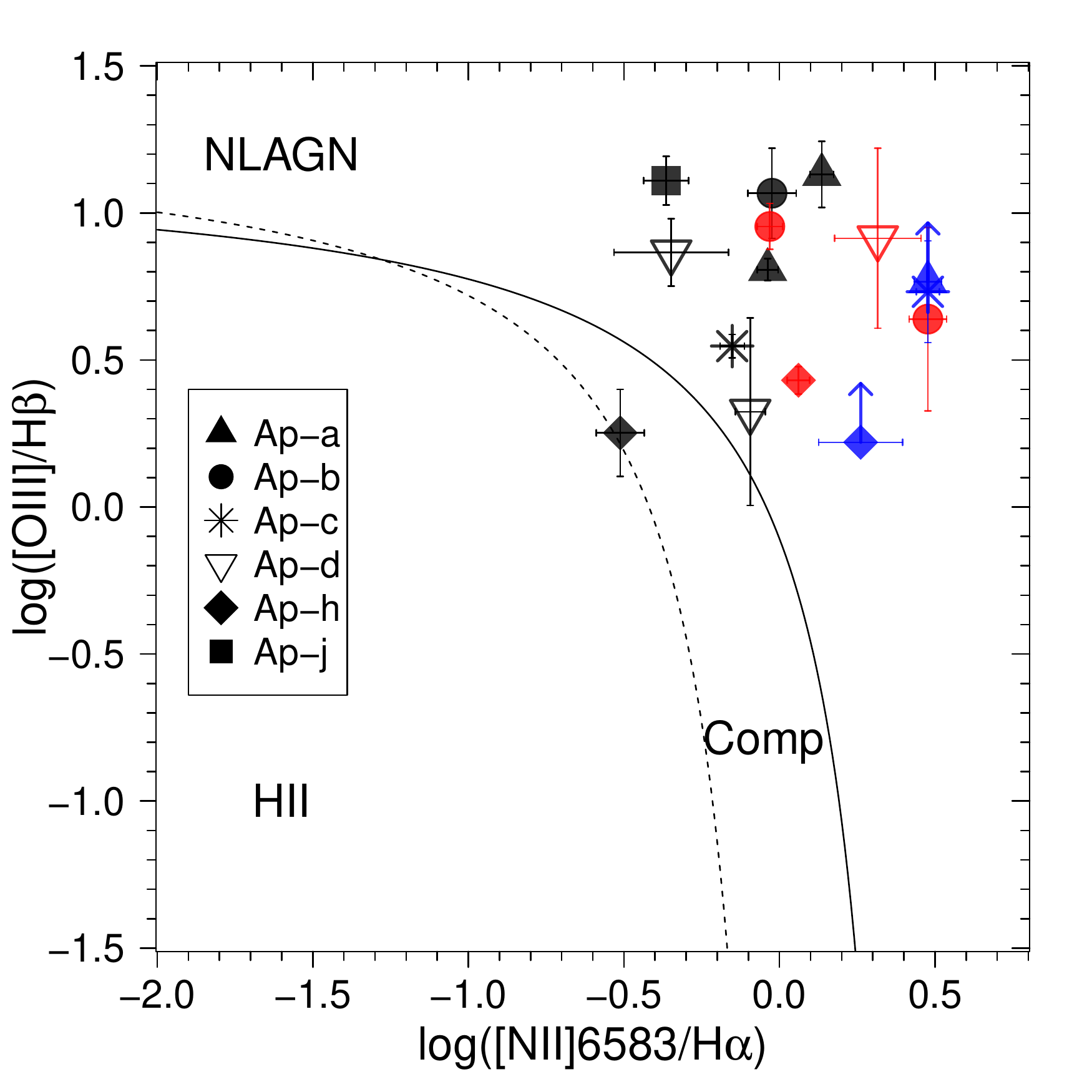}&
\includegraphics[width=0.45\textwidth]{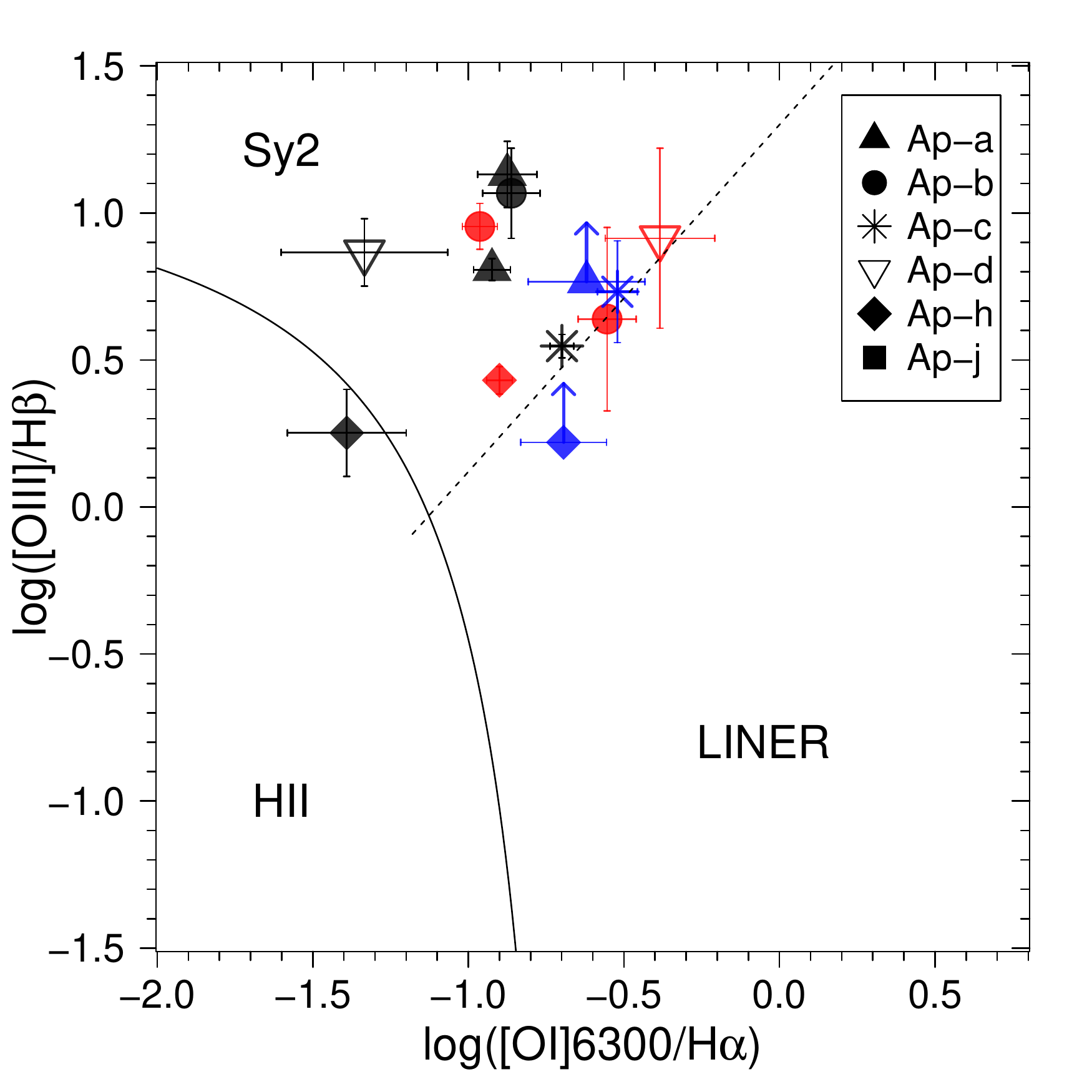}\\
\includegraphics[width=0.45\textwidth]{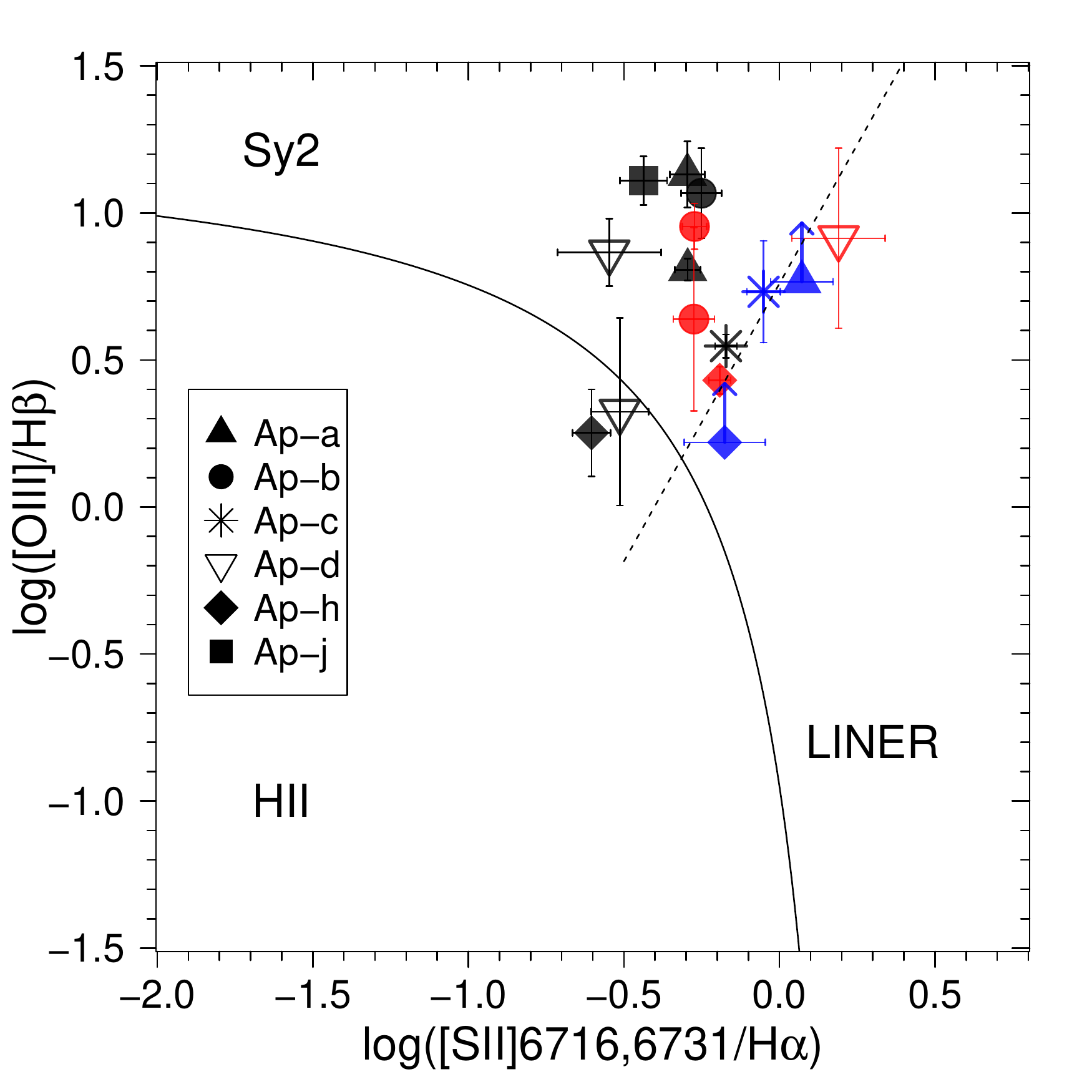}
\end{tabular}
\caption{Same as Figure \ref{BPT-Groves} but for the 6 apertures covering the
  main regions of interest in Mrk273 (see the text for details). The symbols
  correspond to the different apertures and are indicated in the figure, while
  the colors code the FWHM range: black, red and blue correspond to narrow,
  intermediate and broad kinematic components respectively. (A color version of
  this Figure is available in the online journal.)}
\label{BPT-detail}
\end{figure*}

From the results presented above, it is clear that the morphology and kinematics
of the warm ionized gas in Mrk273 are extremely complex. Disturbed emission line
kinematics are observed at almost all locations around the nuclear regions. Our
imaging and spectroscopic results suggest that we are witnessing a variety of
phenomena occurring on different scales in the galaxy. To better understand
these phenomena, we concentrate here in 6 apertures that sample the main regions
of interest and are representative of the diverse kinematic properties
observed. These are: Ap-a, Ap-b, Ap-c, Ap-d, Ap-h and Ap-j (the modeling results
for the H$\alpha$+[NII] complex for these 6 apertures are shown in Figure
\ref{Ha-profiles}). Figure \ref{BPT-detail} shows the line ratios obtained for
these particular apertures, while table \ref{BPT-table} shows the optical
spectral type derived from each line ratio diagram, and the adopted
classification for each extraction aperture.

\subsubsection{The nuclear superbubble and the Outflow-West and Outflow-East structures }

In terms of the kinematics, our results are, in general, consistent with those
of \cite{Colina99} and RV13, i.e. the kinematics in the central region of the
galaxy can be described with a nuclear superbubble oriented N-S and a less
collimated E-W outflow. However, the \cite{Colina99} IFS observations have
relatively low spatial and spectral resolution and the authors required only two
kinematic components to model the emission lines from the galaxy at all
locations. On the other hand, the RV13 observations represent a significant
improvement in sensitivity, spatial and spectral resolution with respect to
those of \cite{Colina99}, although their FOV does not cover most of the region
with enhanced [OIII] emission to the west, and the entire [OIII] emission to the
east of the nuclear region.

In this context, our new {\it HST} observations sample, at higher spatial
resolution, the entire emission from the galaxy, while the {\it INT}-IDS
observations provide spectroscopic information for the nuclear and extended
structures (albeit with lower spatial resolution). For example, the [OIII]
morphology shown in Figure \ref{HST_frames}, and the modeling results in Figures
\ref{Ha-profiles} and \ref{kin-D} allow us to place better constrain the
properties of the E-W outflow. Our results suggest that Outflow-East and
Outflow-West are part of the same outflow phenomenon, extending $\sim$4kpc from
east to west at PA$\sim$80$^{\circ}$ relative to the RV13 nuclear superbubble.

In addition, covering a wavelength range of $\sim$4000--7200\AA, the {\it
  INT}-IDS observations allow us to study the ionization mechanisms responsible
for the emission at these locations. Figure \ref{BPT-detail} and Table
\ref{BPT-table} show that the line ratios at the location of the S-Bubble are
consistent with Sy2 ionization, while those corresponding to the location of
N-Bubble show a mix between Sy2, LINER and HII ionization in the different
BPT/VO87 diagrams. As seen in Figure \ref{HST_frames}, the S-Bubble is located
closer to the SW nucleus and possibly illuminated by the AGN at that nucleus,
which could explain the different line ratios relative to those of the N-Bubble.

In the case of Outflow-West region, line ratios consistent with Sy2 ionization
are found for all the 3 kinematic components in the three BPT/VO87
diagrams. This result, along with the location of the region relative to the SW
nucleus, clearly suggest that this region is being illuminated by the AGN in the
SW nucleus of the galaxy. For Outflow-East, the line ratios show a mix of HII,
LINER and Sy2-like ionization\footnote{The narrowest component (FWHM =
  193$\pm$29 km s$^{-1}$) makes a negligible contribution to the [OI] emission
  within this extraction aperture (Ap-d). Therefore, this component is not
  required to adequately model the [OI] emission line profile (i.e., if
  included, its properties are entirely unconstrained). Hence, no [OI]/H$\alpha$
  line ratio is shown in the corresponding line ratio diagram in Figure
  \ref{BPT-detail}.}.

\setlength{\extrarowheight}{0.05cm}
\begin{table*}
\centering
\begin{tabular}{lllllll}
\hline
Aperture&Region&Comp&&BPT/VO87&&Class\\ \cline{4-6}
&&&[NII]/H$\alpha$&[OI]/H$\alpha$&[SII]/H$\alpha$&\\
(1)&(2)&(3)&(4)&(5)&(6)&(7)\\
\hline
Ap-a$^{\star}$&Outflow-W   &N1&S2&S2&S2&S2\\
    &            &N2&S2&S2&S2&S2\\
    &            &B &S2&S2&L/S2&S2:\\
\hline
Ap-d&Outflow-E   &N1&S2/H& ...& H/S2& Cp\\
    &            &N2&S2&S2&S2&S2\\
    &            &I &S2&S2/L&L/S2&Cp\\
\hline
Ap-b&Bubble count&N &S2&S2&S2&S2\\
    &            &I1&S2&S2&S2&S2\\
    &            &I2&S2&S2/L&S2&S2:\\
\hline
Ap-h$^{\star}$&N-Bubble&N &H/Comp&H/S2&H&Cp\\
    &            &I &S2&S2&S2/L&S2:\\
    &            &B &S2&L/S2&L/S2&Cp\\
\hline
Ap-c&SE compact  &N &S2&S2/L&S2/L&Cp\\
    &            &B &S2&S2/L&S2/L&Cp\\
\hline
Ap-j&Ext. filaments &N &S2&S2&S2&S2\\
\hline
\end{tabular}
\caption{Line ratios measured for the different apertures/regions and for the
  different kinematic components. Col (1): extraction aperture. Col (2):
  corresponding region covered by the extraction aperture. Col (3): kinematic
  component. Col (4)-(6): the optical spectral type for each of the diagnostic
  diagrams. The symbols are: H = HII galaxies, L = LINER and Sy2 = Seyfert 2.
  Col (7): adopted optical spectral type. Cp indicates a mix between S2 and
  LINER and/or HII spectral types. Following \cite{Yuan10} notation, single
  colons indicate apertures with the same spectral classification in two of the
  diagnostic diagrams. \newline $^{\star}$ For the broad component in these two
  apertures we assumed an H$\alpha$ to H$\beta$ ratio of 3. Therefore, the
  [OIII]/H$\beta$ line ratios calculated are lower limits of the real values. }
\label{BPT-table}
\end{table*}
\setlength{\extrarowheight}{0.cm}

\subsubsection{The SE compact structure}

Coinciding exactly with the location of the so-called SE nuclear component we
find the brightest and most compact (FWHM = 0.14 arcsec or 106 pc, see Section
3.1) condensation of [OIII] emission in the galaxy. The nature of the SE
component remains controversial. It was first identified as a star cluster based
on NICMOS images \citep{Scoville00}. \cite{Bondi05} carried out a study of the
source using milliarsecond resolution, high sensitivy radio observations. These
authors found that, at radio wavelengths, the SE structure has a steep spectral
index ($\alpha \simeq$ 1.4) with no compact and/or flat spectrum feature
\citep{Bondi05}, and is fully resolved at 5mas resolution. However, despite the
quality of their observations, the authors could not determine the nature of
this nuclear component in a decisive way. On the other hand, \cite{U13} carried
out an AO-IFU study of the central kpc of the source. Their results based on the
[Si{\small VI}] emission, and the [Si{\small VI}]/Br$\alpha$ line ratios,
suggest that the dominant ionization mechanisms for SE component is AGN
photoionization, although some contribution from shocks cannot be ruled
out. Considering these results, together with the morphology of the nuclear
region observed in their IFU data, \cite{U13} proposed a scenario in which the
SE component is ionized by an AGN in the N nucleus that is completely obscured
along our line of sight at optical wavelengths. This raises the possibility of
Mrk273 being a dual AGN.

Figure \ref{Ha-profiles} shows the H$\alpha$+[NII] emission lines at the at the
location of the SE component. Two kinematic components are required to provide a
good fit to the emission line profiles at this location: a ``narrow'' component
with FWHM of 431$\pm$11 km s$^{-1}$, and a broad component with FWHM =
1407$\pm$47 km s$^{-1}$ redshifted 125$\pm$28 km s$^{-1}$ relative to the narrow
component. The line ratios for these two kinematic components fall in the
limiting region between LINER and Sy2 ionization, and far from the HII
ionization region, in the three BPT/VO87 diagrams.

To shed light on the nature of the SE compact structure it is necessary to
consider both the results in this paper and those of previous studies at other
wavelengths. In the first place, we further emphasize that its structure is
extremely compact, and no morphological connection between this structure and
the N or SW nucleus is observed in our {\it HST} images. In addition, strong
high ionization emission lines are observed at both optical (this study) and
near-IR wavelengths \citep{U13}. Furthermore, there is-significant continuum
emission at this location, especially in the I, H and K bands
\citep{U13}. Together, these results strongly suggest that the SE component is a
separate nucleus with its own AGN, rather than a patch of ISM illuminated by the
AGN in one of the other nuclei \citep{U13}. In this scenario the enhanced
continuum emission in the red-optical and near-IR bands would represent the
direct light of stars in the remnant of the bulge surrounding the putative
supermassive black hole.

The only caveat associated with this scenario is the lack of a clear or X-ray or
near-IR point source at the location of the SE component
\citep{Iwasawa11}. However, this can be explained if the X-ray AGN is heavily
obscured by circumnuclear material, even at near-IR wavelengths, so that we are
seeing the NLR of the AGN rather than the AGN continuum directly. It is also
possible that the AGN at the SE nucleus has an intrinsically lower luminosity
that the AGN at the SW (and N?) component. Perhaps the NLR and any stellar
component associated with the SE nucleus are less obscured that those of the
other nuclei and therefore, appear stronger at optical wavelengths.

Overall, our results are consistent with the idea that the SE component is a
separate nucleus with a strongly obscured, and perhaps relatively low
luminosity, AGN. However, there are some caveats associated with this scenario,
and further observations are required to help elucidate the nature of this
compact structure.

\subsubsection{The extended filaments}

The most spectacular features observed in the {\it HST}-[OIII] and the {\it
  INT}-H$\alpha$ images of the galaxy are the filaments and clumps of ionized
gas emission extended $\sim$30 arcsec ($\sim$23 kpc) to the east of the nuclear
region. Interestingly, the kinematics at all locations in these filaments
covered by our slits are drastically different from those in the nuclear
regions, with no signs of multiple kinematic components. In fact, a kinematic
model comprising a very narrow component (FWHM $\lsim$ 100 km s$^{-1}$) in
sufficient to model the emission lines from the extended filaments. In addition,
our kinematic results suggest the presence of a positive velocity gradient
across the structure, with velocity shifts increasing from $\Delta$V of
-77$\pm$24 km s$^{-1}$ relative to the rest frame in the inner regions to
114$\pm$27 km s$^{-1}$ in the outer regions.

Regarding the ionization mechanisms, for all locations across the extended
filaments the line ratios fall within the Sy 2 ionization region of the 3
BPT/VO87 diagrams shown in Figure \ref{BPT-detail}. Note that, since only one
kinematic component is required to model the emission lines, the degeneracy
issues do not affect the modeling results in this case. In addition, the
continuum emission associated with this structure is relatively weak. Therefore,
the line ratios measured for the ionized gas emission in the extended filaments
and clumps are extremely reliable.

It is interesting to consider how this extended structure might have formed. In
the first place, it is well known that major galaxy mergers are violent events,
with a duration of up to few Gyrs from the first encounter to the time when the
two nuclei finally coalesce
\citep[e.g.][]{Mihos96,Barnes04,Springel05,Cox08,Johansson09}. Numerical
simulations show that, at various stages during the merger event, the gas rains
down into the central regions of the merging galaxies triggering star formation,
and possibly, AGN activity \citep[e.g.][]{Springel05,Cox06,Hopkins10}. This
generates outflows that, intermittently, blow out material from the central
regions to larger scales \citep[e.g.][]{Hopkins05,Springel05,Cox06}. As the
merger event progresses, the high expansion velocity of the material ejected at
early stages gradually decreases due to gravitational effects. Part of this
material will eventually escape from the potential well of the merging system
into the IGM, while the fraction that remains gravitationally bound is accreted
again onto the central region.

In this context, Mrk273 is a merger in its late stages, where the two (or three)
nuclei are close to coalesce. The [OIII] emission from the extended structure
show a bubble-like morphology that, at first sight, suggests that the gas has
been blown out from the nuclear region. Therefore, it is possible that the
emission in the extended filaments is associated with an outflow induced by
AGN-activity at early stages in the merger. In this scenario, the gas in the
filaments has progressively slowed down and any turbulence has partially
dissipated, which would partially explain the emission line properties at these
locations.

Another possibility is that the extended filaments are related to a more recent
outflow event, with the gas in the filaments rapidly expanding in the galaxy
almost perpendicular to the line of sight. In this case, he quiescent kinematics
of the extended emission would be explained in terms of projection
effects. However, if the ionized gas emission in the extended filaments is
really due to outflows (either induced by past AGN activity or related with a
more recent event), one would expect the emission lines from the extended
filaments to show at least some evidence for broadening, due to turbulence in
the outflow. As we described before, we find no evidence for complex kinematics,
and the ionized gas in the extended filament is extremely quiescent (i.e. very
narrow FWHM). Certainly these results do not favour a scenario in which the gas
in the extended filaments is related to an outflow event.

Furthermore, our continuum image, as well as previous broad-band continuum (or
continuum dominated) images from the ground
\citep[][]{Mazzarella93,Hibbard96,Kim02} and the {\it HST} (GO 10592, PI: Aaron
Evans) show significant continuum emission at the locations of the extended
filaments. Such emission is likely to trace the starlight. However, it seems
unlikely that stars would be able to form in the extreme conditions of an
outflow \citep[but see Tadhunter et al. 2014 and][]{Zubovas14}. Therefore,
considering all these results, we favour a scenario in which the entire ionized
gas emission from the filaments and clumps represents tidal debris left over
from a secondary merger event that is illuminated by one of the AGNs in the
galaxy. This would be consistent with the quiescent kinematics of the extended
filaments, the Sy2 ionization line ratios and also the presence of continuum
emission, due to stars from one of the precursor galaxies than now form part of
the tidal debris. Note that this scenario resembles that described by
\cite{McDowell03} for Arp220. In that paper, the authors found that ``lobes'' of
extended, faint, edge-brightened ionized gas emission were likely related to the
merger dynamics rather than the result of an outflow. Therefore, it is possible
that such low surface brightness features are common among these types of merger
remnants, but have so far been difficult to detect.


In terms of the relationship between the extended filaments and the nuclear
structures, the morphology of the emission from the filaments closer to the
nuclear region suggest that they emerge from the N nucleus. This is an
interesting result since, as we mentioned before, the line ratios at all
location in the filaments are consistent with Sy2 ionization. However, to date,
there is no direct evidence for the presence of an AGN in that nucleus. At this
stage it is worth remembering that the N nucleus is embedded in a massive
rotating disk of molecular gas at PA = 70$^{\circ}$--90$^{\circ}$
\citep[][]{Downes98,Cole99}. Therefore, it is possible that the AGN in that
nucleus is too heavily obscured to be detected in the current optical and
infrared observations of the source. In addition, it is unlikely that any
putative AGN within that disk illuminates the entire [OIII] nebula, as the disk
would absorb the radiation along the major axis. In this context, it is possible
that the extended filaments are tidal debris from the galaxy that hosted the N
nucleus, but are illuminated by the AGN at the SW (or the SE) component. A
second possibility is that he AGN at the N nucleus has temporarily switched off
so that the extended emission from the filaments represents a ``light echo'',
reflecting a previous phase of AGN activity.

Clearly, further observations are required to investigate the origin of the
extended filamentary structure and its relationship with the nuclear structures.
For example, large FOV, deep IFS observations covering the full extent of the
filaments would help to distinguish between the possible scenarios described in
this section for the origin of this structure. In addition, they would be
extremely useful to investigate in detail which of the nuclear structures is
directly related to the extended filaments.

\section{Summary}

We have used ACs/HST and INT/WFC observations to carry out an optical imaging
and spectroscopic study of the warm, ionized gas in the nuclear and extended
regions of Mrk273. The results from this study show that the emission line
kinematics in this galaxy are extremely complex, with multiple outflows
occurring in the nuclear regions of the galaxy out to a radial distance of 4kpc
from the N nucleus. In addition, we find emission line ratios that are
consistent with AGN photoionization as the dominant ionizing mechanism at most
locations in the galaxy sampled by our slit. Shocks may also be present, but our
results suggest that are not the dominant source of ionization. To gain a better
idea of the physical processes ongoing in Mrk273 we have concentrated on 6
extraction apertures that sample the main regions of interest, and have
investigated in detail their kinematics and ionization mechanisms. The results
can be summarised as follows.

\begin{itemize}

\item {\bf The nuclear superbubbles}: the presence of nuclear superbubbles
  oriented N-S was reported by RV13. Our {\it HST} [OIII] image of the galaxy
  shows a ``U-shaped'' structure to the north of the nuclear region that traces
  well the northern bubble (N Bubble). The Southern bubble (S Bubble) is not
  visible in our ACS image; this can be explained in terms of reddening
  effects. The kinematics measured at the locations of the bubbles are
  consistent with those measured by RV13, and are the most extreme in the
  galaxy. The line ratios indicate a mix between HII, LINER and Sy2 mechanisms
  for N-Bubble, while S-Bubble has line ratios typical of Sy2 ionization for all
  the kinematic components. These results are consistent with the nuclear
  superbubbles emerging from the N nucleus of the galaxy. The proximity of
  S-Bubble to the SW nucleus could explain the different line ratios between the
  two sides of the bubbles. Of course, another possibility is that the line
  ratios in S-Bubble are directly related to an AGN buried the N nucleus.

\item {\bf Outflow-West and Outflow-East}: Three kinematic components are
  required to model the emission from both Outflow-West and Outflow-East. In the
  case of Outflow-West, emission line ratios consistent with Sy2 ionization are
  found for the 3 kinematic components in the three BPT/VO87 diagrams. These
  results, along with the location of the Outflow-West in the galaxy clearly
  suggest that this region is being illuminated by the AGN in the SW
  nucleus. For Outflow-East the line ratios show a mix between HII, LINER and
  dominant Sy2 ionization. Our results suggest that Outflow-East is related to
  Outflow-West and that the two structures correspond to a nuclear outflow
  expanding from west to east almost perpendicular to the nuclear superbubbles.

\item{\bf The SE compact structure}: coinciding with the location of the so
  called SE nuclear component, this is the brightest and most compact structure
  observed in the {\it HST} [OIII] image of the galaxy. The compact nature of
  this structure, the high ionization state derived from our optical study and
  the near-IR study of \cite{U13}, and the presence of significant continuum
  emission at this location in the I, H and K bands \citep{U13} together
  strongly suggests that the SE component is a separate nucleus with its own AGN
  rather than a clump of ionized gas illuminated by one of the other AGNs, as
  proposed by \cite{U13}.

\item{\bf The extended filaments}: the ACS-[OIII] and {\it INT}-H$\alpha$ images
  of the galaxy show show a spectacular system of filaments and knots of
  emission extending $\sim$23 kpc to the east of the nuclear region. A kinematic
  model comprising a a single, narrow Gaussian component provides a good fit to
  the emission lines, and the lines ratios are consistent with Sy2 ionization,
  at all locations in the filament system. These results suggest that the gas
  emission in the filaments is associated with tidal debris left over from a
  secondary merger event, with the filaments ionized by an AGN in the nuclear
  regions of the galaxy. Indeed, it is possible that that Outflow-West,
  Outflow-East and the extended filaments represent ionization cones emerging
  from one of the nuclei.

\end{itemize}

Mrk273 is a nearby ULIRG with a variety of interesting phenomena occurring at
all scales in the galaxy. Therefore, it offers a perfect opportunity to
investigate the physical processes that govern the relationship between the
nuclear activity and the surrounding ISM in gas-rich mergers. The new {\it HST}
images reveal a very disturbed morphology in the nuclear region of the galaxy,
and a spectacular filamentary structure extending $\sim$23kpc to the east of the
galaxy whose origin remains uncertain. In addition, we have identified the
brightest and the most compact structure in [OIII] emission with the SE
component. Our results suggest that this component is a nucleus itself, with its
own AGN. If this result is confirmed, as well as those of \cite{U13} which
suggest the presence of an AGN in the N disk, three AGN could potentially be
co-existing in the nuclear region of Mrk273.

Overall, these results demonstrate the potential of deep observations of the
extended emission line gas for unravelling the complexity of ULIRGs, determining
the nature of triggering mergers, and the gauging the importance of feedback
effects associated with their AGN.


\section*{Acknowledgments}

JRZ and CRA acknowledge financial support from the spanish grant
AYA2010\_21887-C04-04. The Isaac Newton Telescope is operated on the island of
La Palma by the Isaac Newton Group in the Spanish Observatorio del Roque de los
Muchachos of the Instituto de Astrofisica de Canarias. JRZ would like to thank
Manuel D\'iaz Alfaro and Javier Piqueras L\'opez for the help with the R and IDL
codes used in this paper. This research is supported in part by STFC grant
St/J001589/1 and a Marie Curie IEF within the 7th European community framework
programme (PIEF-GA-2012-327934).

\bibliographystyle{aa} \bibliography{JRZRefs}

\end{document}